\documentclass[aps,pre,showpacs,twocolumn]{revtex4}
\usepackage{bm}
\usepackage{graphicx}
\bibstyle{approve.bib}
\usepackage{amssymb}
\usepackage{amsmath}
\usepackage{esint}
\usepackage{epstopdf}
\usepackage{color}
\usepackage{gensymb}
\usepackage{mathtools}
\usepackage{suffix}
\usepackage{fancyhdr}

\newcommand{\be}{\begin{equation}}
\newcommand{\ee}{\end{equation}}
\newcommand{\bea}{\begin{eqnarray}}
\newcommand{\eea}{\end{eqnarray}}
\newcommand{\ph}{\mathbf{\Phi}}
\newcommand{\lan}{\left\langle}
\newcommand{\ran}{\right\rangle}
\newcommand{\br}{\mathbf{r}}
\newcommand{\bR}{\mathbf{R}}

\newcommand{\hk}{\hat{k}}

\newcommand{\bq}{\mathbf{q}}

\newcommand{\e}{\varepsilon}

\newcommand{\tv}{\tilde{v}}
\newcommand{\tq}{\tilde{Q}}

\newcommand{\may}{\tilde{h}_{ij}(q)}

\newcommand{\tchi}{\tilde{\chi}}
\newcommand{\tG}{\tilde{G}}
\newcommand{\tgg}{\tilde{{\mathcal G}}}

\newcommand{\tep}{\tilde{\varepsilon}}

\newcommand{\ce}{_{\rm c}}
\newcommand{\G}{_{\rm G}}
\newcommand{\el}{_{\rm el}}
\newcommand{\h}{_{\rm h}}
\newcommand{\lo}{_{\rm l}}
\newcommand{\n}{_{\rm n}}

\newcommand{\s}{_{\rm s}}

\newcommand{\hn}{\hat{n}}

\newcommand{\ew}{\varepsilon_{\rm s}}

\begin{document}

\title{Unified microscopic theory of equilibrium thermodynamics and ion association in aqueous and non-aqueous electrolytes with explicit hard-core size}

\author{Sahin Buyukdagli}
\address{Department of Physics, Bilkent University, Ankara 06800, Turkey}

\begin{abstract}

\noindent{\bf ABSTRACT}: Within the framework of a functional integral formalism incorporating ionic charge and hard-core (HC) interactions on an equal footing, we formulate a unified theory of equilibrium thermodynamics and ion association in charged solutions. Via comparison with recent Monte-Carlo (MC) simulation results (J. Forsman et al., PCCP {\bf 26}, 19921 (2024)), it is shown that our approach is able to predict with quantitative precision the pair distributions of monovalent ions with the typical hydrated sizes $d=3.0$ {\AA} and $4.0$ {\AA} up to the molar concentration $n_i\approx2.0$ M. Moreover, comparison with additional simulation data from the literature indicates that within the characteristic regime of ionic packing fraction $\eta\lesssim0.1$, the formalism can accurately account for the ion size dependence of the excess energy and pressure from $d=14.3$ {\AA} down to $d=1.6$ {\AA}. Via the adjustment of the hydration  radius, the theory can also reproduce the non-monotonic salt dependence of the experimentally measured osmotic coefficients of various aqueous and non-aqueous solutions. In accordance with AFM experiments involving non-aqueous electrolytes, the underlying sharp competition between the opposite charge attraction and the excluded volume constraint is shown to limit the occurrence of substantial ionic pair formation to the submolar concentration regime $n_i\lesssim50$ mM; at larger concentrations, HC repulsion hinders ion association and results in the quasi-saturation of the pair fraction curves.

\end{abstract}

\date{\today}
\maketitle   

\section{Introduction}

Polar fluids play a central role in the coordination of the biochemical processes enabling the functioning of living systems within a self-sustained cycle. Owing to the omnipresence of free charges in nature, the mechanism of hydration occupies a critical place in the regulation of the competition between electrostatic attraction and HC repulsion mediating the constant interaction of solute particles and biomolecules. Hence, the accurate characterization of this competition with respect to the solvent polarity is essential for the control of numerous biophysical and artificial processes ranging from active transport~\cite{Bont,biomatter} and viral infection~\cite{gn1} to electrokinetic energy conversion~\cite{Boc1,Boc2} and water nanofiltration~\cite{Yar,Szymczyk}.

The Debye-H\"{u}ckel (DH) theory formulated a century ago~\cite{DH} has been a valuable theoretical approach enabling the characterization of the equilibrium thermodynamics of charged solutions. Neverthless, as an electrostatic weak-coupling (WC) theory neglecting the ionic HC size, the accuracy of the DH formalism is strictly limited to dilute monovalent electrolytes~\cite{Buyuk2024}. As a result, the study of condensed solutions and strongly charged systems had to rely on MC simulations~\cite{Val80,Svensson,NetzMC,NetzMC2,LevinMC} as well as numerically involved theoretical approaches such as density functional and integral equation theories~\cite{Blum,Henderson,Boda,Hoye,Hansen,Roji}.

On the side of the analytically tractable approaches, the thermodynamics of charged liquids can be alternatively formulated via the partition function of an effective single charge coupled to a fluctuating electrostatic potential dressed by many-body interactions. Upon its formulation by Kholodenko et al.~\cite{Kholodenko1,Kholodenko2}, it has been realized that this compact field theoretic formalism is practical for the incorporation of various relevant complications omnipresent in real electrolytes, such as explicit solvent molecules~\cite{PRE2013,PRE2022}, the inner solute structure~\cite{PRE2023} and polarizability~\cite{podgornik2}, and charged biomolecules with conformational degrees of freedom~\cite{Cherstvy2011}. Due to its mathematical transparency, the functional integral form of the electrostatic partition function is also suitable for analytical treatment by various approximation schemes, such as systematic WC~\cite{NetzLoop} and strong-coupling perturbation techniques~\cite{NetzSC}, and variational methods~\cite{netzvar,HatloVar0,HatloSoft,HatloEpl,Buyuk2012,Buyuk2020}. 

Due to the emergence of sizeable HC effects at submolar salt concentrations, the aforementioned functional integral approaches neglecting the HC ion size do not give access to molar concentrations commonly encountered in aqueous electrolytes~\cite{Book}. With the aim to overcome this limitation, we have recently developed a cumulant-corrected DH (CCDH) approach explicitly incorporating HC interactions into the field theoretic formulation of charged solutions~\cite{Buyuk2024}. This upgrade enabled the accurate evaluation of the HC-dominated quantities such as the osmotic coefficient and the activity coefficient up to the molar concentration regime.  However, the gaussian closure of the Schwinger-Dyson (SD) identities at the basis of the CCDH theory leads to the WC treatment of the strongly coupled charge interactions occurring at short interionic distances. This decreases in turn the accuracy of the CCDH approach in predicting the electrostatically governed thermodynamic functions such as the excess energy and the screening length characterizing the spatial range of charge interactions. 

In order to remedy this deficiency, in Ref.~\cite{Soft2024}, we have incorporated into the CCDH formalism a mathematically structured and generalized version of the splitting approach originally introduced in Refs.~\cite{sp,Chen,Santangelo,HatloSoft,HatloEpl,LueSoft}. Our variational splitting technique enabling the asymmetric treatment of distinct interaction scales allows to incorporate the long-range interactions within the WC gaussian approach and the short-range interactions via the virial approximation. Owing to this mixed approximation strategy, the resulting self-consistent DH (SCDH) formalism avoids the WC treatment of the strongly coupled short-range electrostatic and HC interactions. 

In the present work, we introduce a computationally enhanced version of the SCDH theory upgraded by the analytical evaluation of the gaussian-level correlation function. By reducing the number of numerical Fourier Transforms (FTs) involved in the formalism, this upgrade enhances significantly the execution speed and also improves the numerical convergence of our computational approach. The formalism is tested via comparison with a large set of simulation results and experimental data, and it is used for the characterization of the ionic pair formation in electrolytes of diverse polarity.

Our article is organized as follows. Section~\ref{th} introduces the electrolyte model, the SCDH formalism, and an ionic association model enabling the evaluation of the ionic pair fractions from the partition function. In Sec.~\ref{compMC}, via the systematic comparison of the radial distributions and the thermodynamic functions obtained from the SCDH approach with a large variety of MC simulation data~\cite{NetzMC,Val80,Svensson,Val90,Ork94} including the recent simulation results of Ref.~\cite{Forsman}, we identify the validity regime of the present formalism in terms of salt concentration and ion size. Then, in Sec.~\ref{expe}, we confront the predictions of our theory with the experimental osmotic coefficient data of various aqueous and non-aqueous electrolytes. Finally, Sec.~\ref{ass} is devoted to the characterization of the molecular mechanisms driving the ion association phenomenon~\cite{andelman,safran} responsible for the underscreening of macromolecular interactions in non-aqueous electrolytes~\cite{ExpSc,ExpSc2,ExpAlc1}.

\section{Theory}
\label{th}

In this section, we introduce the electrolyte model and derive the functional integral representation of its partition function. Then, we explain the derivation of the SCDH formalism including its upgraded features such as the analytical calculation of the gaussian-level correlation function that significantly speeds up the numerical computations underlying our formalism, an exact identity relating the ionic pair distributions and the electrostatic two-point correlation functions, and the ion association model enabling the calculation of the ionic pair fractions.

\subsection{Electrolyte Model and Partition Function}

The liquid is composed of $p$ ion species. Each ion of the species $i$ with valency $q_i$ and concentration $n_i$ is placed at the center of a HC sphere with diameter $d$. The ions are hydrated by a solvent of temperature $T$ and dielectric permittivity $\ew\e_0$, where $\e_0$ and $\ew$ stand for the vacuum permittivity and the relative dielectric constant, respectively. Moreover, two ions separated by the distance $r$ interact via (i) the HC potential $v\h(r)$ defined as
\be
\label{eq0}
e^{-v\h(r)}=\theta(r-d),
\ee
where $\theta(x)$ is the Heaviside theta function~\cite{math}, and (ii) the electrostatic Coulomb potential $v\ce(r)=\ell_{\rm B}/r$ corresponding to the inverse of the bulk Coulomb operator
\be\label{eq0II}
v\ce^{-1}(\br,\br')=-\frac{1}{4\pi\ell_{\rm B}}\nabla^2\delta^3(\br-\br'),
\ee
with the Bjerrum length $\ell_{\rm B}=e^2/(4\pi\ew\e_0 k_{\rm B}T)$ including the electron charge $e$ and the Boltzmann constant $k_{\rm B}$.

The grand canonical (GC) partition function of the electrolyte corresponding to the trace of the Boltzmann distribution function over the fluctuating particle numbers $N_i$ and center of mass positions $\br_{jk}$ reads
\bea
\label{eq1}
Z\G=\prod_{i=1}^p\sum_{N_i=0}^\infty\frac{\lambda_i^{N_i}}{N_i!}\prod_{j=1}^p\prod_{k=1}^{N_j}\int\mathrm{d}^3\br_{jk}e^{-\beta(E\ce+E\h+E\n)},
\eea
where $\lambda_i$ stands for the ion fugacity. Eq.~(\ref{eq1}) includes as well the pairwise coupling energies 
\bea
\label{eq2}
\beta E_\alpha=\frac{1}{2}\int\mathrm{d}^3\br\mathrm{d}^3\br'\hn_\alpha(\br)v_\alpha(\br,\br')\hn_\alpha(\br')
\eea
for Coulombic ($\alpha={\rm c}$) and HC interactions ($\alpha={\rm h}$), and  the one-body energy component 
\be
\label{eq3}
E\n=\sum_{i=1}^p\int\mathrm{d}^3\br\;w_i(\br)\hn_i(\br)-\sum_{i=1}^pN_i\epsilon_i
\ee
including the steric ion potential $w_i(r)$ to be used for the computation of the average ion densities, and the ionic self-energy $\epsilon_i=\left[q_i^2v\ce(0)+v\h(0)\right]/2$ subtracted from the total energy. We finally note that Eqs.~(\ref{eq2})-(\ref{eq3}) involve the total number and charge density operators
\be\label{eq4}
\hn\h(\br)=\sum_{i=1}^p\hn_i(\br);\hspace{5mm}\hn\ce(\br)=\sum_{i=1}^pq_i\hn_i(\br)+F\ce(\br)
\ee
defined in terms of the number density operator
\be\label{eq5}
\hn_i(\br)=\sum_{j=1}^{N_i}\delta^3(\br-\br_{ij}),
\ee
and the fixed charge density $F\ce(r)$.

Following the splitting scheme originally introduced in Refs.~\cite{Santangelo,HatloEpl,HatloSoft}, we separate now the Coulomb interaction potential into a short-range component $v\s(\br)$ and a long-range component $v\lo(\br)$ whose functional forms will be specified below,
\be
\label{eq6}
v\ce(\br)=v\s(\br)+v_{\rm l}(\br).
\ee
In Eq.~(\ref{eq2}) for $\alpha=c$, this splitting gives rise to two electrostatic pairwise interaction terms. Thus, introducing in Eq.~(\ref{eq1}) three separate Hubbard-Stratonovich transformations of the form
\bea\label{eq9}
\hspace{-5mm}&&e^{-\frac{1}{2}\int\mathrm{d}^3\br\mathrm{d}^3\br'\hn_\gamma(\br)v_\gamma(\br-\br')\hn_\gamma(\br')}\\
\hspace{-5mm}&&=\int\mathcal{D}\phi_\gamma\;e^{-\frac{1}{2}\int\mathrm{d}^3\br\mathrm{d}^3\br'\phi_\gamma(\br)v^{-1}_\gamma(\br-\br')\phi_\gamma(\br')}e^{i\int\mathrm{d}^3\br \;\hn_\gamma(\br)\phi_\gamma(\br)}\nonumber
\eea
for the corresponding short-range ($\gamma={\rm s}$) and long-range ($\gamma={\rm l}$) electrostatic interactions, and the HC interactions ($\gamma={\rm h}$), the GC partition function can be finally recast as the following functional integral
\be
\label{eq10}
Z_{\rm G}=\int\frac{\mathcal{D}\ph}{\sqrt{{\rm det}\left[v\s v\lo v\h\right]}}\;e^{-\beta H[\ph]}.
\ee
In Eq.~(\ref{eq10}), we introduced the shorthand vector notations for the fluctuating potentials $\ph=(\phi\s,\phi\lo,\phi\h)$ and the functional integration measure $\mathcal{D}\ph=\mathcal{D}\phi\s\mathcal{D}\phi\lo\mathcal{D}\phi\h$, and defined the Hamiltonian functional 
\bea\label{eq11}
\beta H[\ph]&=&\sum_{\gamma=\{{\rm s,l,h}\}}\int\frac{\mathrm{d}^3\br\mathrm{d}^3\br'}{2}\phi_\gamma(\br)v_\gamma^{-1}(\br,\br')\phi_\gamma(\br')\\
&&-\sum_{i=1}^p\lambda_i \int\mathrm{d}^3\br\;\hk_i(\br)-i\int\mathrm{d}^3\br\left[\phi\lo+\phi\s\right]_\br F\ce(\br)\nonumber
\eea
including the fluctuating ion density function
\be\label{eq12}
\hk_i(\br)=e^{\epsilon_i-w_i(\br)+i\phi\h(\br)+iq_i\left[\phi\s(\br)+\phi\lo(\br)\right]}.
\ee

In the remainder,  the ionic steric potential $w_i(\br)$ and the test charge density $F\ce(\br)$ will be used exclusively for the derivation of the average ion densities, the pair distribution functions, and the electrostatic two-point correlation function (2PCF). Thus, unless stated otherwise, these functions will be set to zero, i.e. $w_i(\br)=F\ce(\br)=0$.

\subsection{Splitting Scheme}

In this article, the splitting in Eq.~(\ref{eq6}) is specified by choosing the inverse of the long-range potential as the following operator originally introduced in Ref.~\cite{HatloSoft},
\be\label{eq13}
v^{-1}\lo(\br,\br')=\left(1-\sigma^2\nabla^2+\sigma^4\nabla^4\right)v\ce^{-1}(\br,\br').
\ee
In Eq.~(\ref{eq13}), the arbitrary length $\sigma$ separating the short- and long-wavelength interactions will be determined variationally. In the present work, the FT and the inverse FT of the general function $f(\br)$ are defined as $\tilde{f}(q)=\int\mathrm{d}^3\br f(\br)e^{-i\bq\cdot\br}$ and $f(\br)=(2\pi)^{-3}\int\mathrm{d}^3\bq \tilde{f}(q)e^{i\bq\cdot\br}$, respectively. Thus, Fourier-transforming Eqs.~(\ref{eq0II}) and~(\ref{eq13}), and using the constraint~(\ref{eq6}), the short- and long-range potential components follow in reciprocal space in the form
\bea
\label{eq14}
\hspace{-7mm}\tv_{\rm s}(q)&=&\tv\ce(q)\frac{\sigma^2q^2+\sigma^4q^4}{1+\sigma^2q^2+\sigma^4q^4};\hspace{3mm}\tv\ce(q)=\frac{4\pi\ell_{\rm B}}{q^2};\\
\label{eq15}
\hspace{-7mm}\tv_{\rm l}(q)&=&\tv\ce(q)\left(1+\sigma^2q^2+\sigma^4q^4\right)^{-1}.
\eea
Calculating now the inverse FT of Eqs.~(\ref{eq14})-(\ref{eq15}), these potential components finally follow in real space as
\bea
\label{eq16}
v_{\rm s}(r)&=&\frac{\ell_{\rm B}}{r}\left\{\cos\left(\frac{r}{2\sigma}\right)+\frac{1}{\sqrt{3}}\sin\left(\frac{r}{2\sigma}\right)\right\}e^{-\frac{\sqrt{3}r}{2\sigma}};\\
\label{eq17}
v_{\rm l}(r)&=&\frac{\ell_{\rm B}}{r}-v_{\rm s}(r).
\eea

\subsection{Variational determination of the splitting length $\sigma$}

The evaluation of the parameter $\sigma$ in Eqs.~(\ref{eq14})-(\ref{eq15}) will be based on the invariance of the partition function~(\ref{eq1}) and the grand potential $\Omega_{\rm G}=-k_{\rm B}T\ln Z_{\rm G}$ under the variation of this characteristic length. Thus, expressing the corresponding condition $\partial_\sigma\Omega_{\rm G}=0$ via the functional integral representation~(\ref{eq10}) of the partition function, one obtains the formally exact variational identity
\be\label{eq18}
\sum_{\gamma=\{{\rm s,l}\}}\hspace{-.5mm}\int\hspace{-1mm}\mathrm{d}^3\br\mathrm{d}^3\br'\left[\partial_{\sigma}v_\gamma^{-1}(\br,\br')\right]\hspace{-1mm}\left\{G_\gamma(\br,\br')-v_\gamma(\br,\br')\right\}=0.
\ee
In Eq.~(\ref{eq18}), the 2PCFs 
\be
\label{eq19}
G_\gamma(\br,\br')=\lan\phi_\gamma(\br)\phi_\gamma(\br')\ran
\ee
associated with the short-range ($\gamma={\rm s}$) and long-range ($\gamma={\rm l}$) interactions involve the functional average defined for a general functional $F[\ph]$ as
\be\label{eq20}
\lan F[\ph]\ran=\frac{1}{Z_{\rm G}}\int\mathcal{D}\ph\;e^{-\beta H[\ph]}F[\ph].
\ee
Finally, exploiting the translational invariance implying $v_\gamma(\br,\br')=v_\gamma(\br-\br')$ and $G_\gamma(\br,\br')=G_\gamma(\br-\br')$, the FT of the variational equation~(\ref{eq18}) follows in the simpler form 
\bea
\label{eq21}
\sum_{\gamma=\{{\rm s,l}\}}\int_0^\infty\mathrm{d}qq^2\left\{\tG_\gamma(q)-\tv_\gamma(q)\right\}\partial_\sigma\tv^{-1}_\gamma(q)=0.
\eea

\subsection{Ion density and pair distribution function}

For the evaluation of the thermodynamic functions investigated in the present work, one needs to connect the ion fugacities to the ion densities and the pair distribution functions. The average ion density corresponds to the GC average of the density operator~(\ref{eq5}), i.e. $n_i=\lan\hn_i(\br)\ran_{\rm G}$. Via Eqs.~(\ref{eq1}) and~(\ref{eq3}), this can be expressed as $n_i=-\delta\ln Z_{\rm G}/\delta w_i(\br)$. Plugging into this identity the functional integral form~(\ref{eq10}) of the partition function, the ion concentration follows in terms of the functional average of the density function~(\ref{eq12}) as
\be
\label{eq25}
n_i=\lambda_i\lan\hk_i(\br)\ran.
\ee

The pair distribution function of the ion species $i$ and $j$ is defined as
\be
\label{eq26}
g_{ij}(\br,\br')\hspace{-0.5mm}=\hspace{-0.5mm}\lan\sum_{k=1}^{N_i}\delta^3(\br-\br_{ik})\hspace{-0.5mm}\sum_{l=1}^{N_j}\delta^3(\br'-\br_{jl})\frac{1-\delta_{ij}\delta_{kl}}{n_in_j}\ran_{\rm G},
\ee
where the Kronecker delta symbols $\delta_{ij}$ subtract the self-interactions. Using the definition of the density operator~(\ref{eq5}) and the average ion density $n_i=\lan\hn_i(\br)\ran_{\rm G}$, Eq.~(\ref{eq26}) can be reduced to
\be\label{eq27}
n_in_jg_{ij}(\br,\br')=\lan\hn_i(\br)\hn_j(\br')\ran_{\rm G}-n_i\delta_{ij}\delta^3(\br-\br').
\ee
Via the identity~(\ref{eq3}), one can now express Eq.~(\ref{eq27}) in terms of the partition function~(\ref{eq1}) as
\be
\label{eq28}
n_in_jg_{ij}(\br,\br')=\frac{1}{Z_{\rm G}}\frac{\delta^2Z_{\rm G}}{\delta w_i(\br)\delta w_j(\br')}-n_i\delta_{ij}\delta(\br-\br').
\ee
Plugging into Eq.~(\ref{eq28}) the functional integral representation of the partition function~(\ref{eq10}), one finally obtains
\be
\label{eq30}
g_{ij}(\br,\br')=\frac{\lambda_i\lambda_j}{n_in_j}\lan k_i(\br)k_j(\br')\ran.
\ee

\subsection{Derivation of the Schwinger-Dyson identities}

The derivation of the thermodynamic identities obtained in the remainder will be based on the electrostatic SD equations relating the fluctuating potential averages to the physical parameters of the electrolyte~\cite{Soft2024}. The calculation of the SD identities requires the definition of the following functional integral~\cite{justin},
\be\label{eq22}
J=\int\mathcal{D}\ph\;e^{-\beta H[\ph]}F[\ph].
\ee
Introducing in Eq.~(\ref{eq22}) the infinitesimal potential shift $\phi_\gamma(\br)\to\phi_\gamma(\br)+\delta \phi_\gamma(\br)$ for $\gamma=\{{\rm s,l}\}$, the resulting variation of this integral follows at the order $O\left[\delta\phi_\gamma(\br)\right]$ as
\bea
\label{eq23}
&&\hspace{-3mm}\delta J=\int\mathrm{d}^3\br\delta \phi_\gamma(\br)\int\mathcal{D}\ph\;e^{-\beta H[\ph]}\\
&&\hspace{3.cm}\times\left\{F[\ph]\frac{\delta\left(\beta H[\ph]\right)}{\delta\phi_\gamma(\br)}-\frac{\delta F[\ph]}{\delta\phi_\gamma(\br)}\right\}.\nonumber
\eea
Accounting now for the invariance of the integral~(\ref{eq22}) under this potential shift that could be removed via a change of the functional integration variable, i.e. setting Eq.~(\ref{eq23}) to zero, one obtains the general SD identities
\be
\label{eq24}
\lan \frac{\delta F[\ph]}{\delta\phi_\gamma(\br)}\ran=\lan F[\ph]\frac{\delta\left(\beta H[\ph]\right)}{\delta\phi_\gamma(\br)}\ran.
\ee

\subsection{Global electroneutrality condition}

In order to derive the global electroneutrality constraint, in the SD Eq.~(\ref{eq24}), we set $\gamma=l$ and $F[\ph]=1$ to obtain $\lan\delta H[\ph]/\delta\phi\lo(\br)\ran=0$. Then, substituting the Hamiltonian functional~(\ref{eq11}) into the latter equality, and using Eq.~(\ref{eq25}), one gets 
\be
\label{eq31}
\tv^{-1}\lo(0)\bar{\phi}\lo=i\sum_{i=1}^pn_iq_i,
\ee
where we took into account the uniformity of the external potential $\bar{\phi}\lo=\lan\phi\lo(\br)\ran$ originating from the translational invariance in the bulk solution. Finally, noting that the infrared (IR) limit of the Fourier-transformed inverse of the long-range potential~(\ref{eq15}) vanishes, i.e. $\tv^{-1}\lo(0)=0$, Eq.~(\ref{eq31}) yields the global electroneutrality condition
\be
\label{eq32}
\sum_{i=1}^pn_iq_i=0.
\ee

\subsection{Two-point correlation function}

Within the present formalism, the moment conditions and the screening parameter will be extracted from the net electrostatic 2PCF. In order to derive the latter, we choose the fixed charge sources in Eq.~(\ref{eq11}) as $n$ point-like test charges of valency $C_m$ and position $\bR_m$,
\be
\label{eq33}
F\ce(\br)=\sum_{m=1}^nC_m\delta^3(\br-\bR_m).
\ee
The 2PCF characterizing the electrostatic coupling of the test charges $C_i$ and $C_j$ corresponds to the susceptibility of the potential $\bar\phi(\bR_i)=\partial_{C_i}\left(\beta\Omega_{\rm G}\right)$ induced by the charge $C_i$ to the presence of the charge $C_j$, i.e. $\mathcal G(\bR_i,\bR_j)=\left\{\partial_{C_j}\bar{\phi}(\bR_i)-\bar{\phi}(\bR_i)\bar{\phi}(\bR_j)\right\}_{C_{\bf m}=0}$, or
\be\label{eq34}
\mathcal G(\bR_i,\bR_j)=\left\{\frac{\partial^2\left(\beta\Omega_{\rm G}\right)}{\partial C_i\partial C_j}-\frac{\partial\left(\beta\Omega_{\rm G}\right)}{\partial C_i}\frac{\partial\left(\beta\Omega_{\rm G}\right)}{\partial C_j}\right\}_{C_{\bf m}=0}.
\ee
Evaluating Eq.~(\ref{eq34}) with the grand potential $\Omega_{\rm G}=-k_{\rm B}T\ln Z_{\rm G}$ expressed in terms of the functional integral representation~(\ref{eq10}) of the partition function, one obtains
\be
\label{eq35}
\mathcal G(\br,\br')=\hspace{-2mm}\sum_{\gamma=\{{\rm s,l}\}}\hspace{-2mm}\lan\phi_\gamma(\br)\phi_\gamma(\br')\ran+\lan\phi\s(\br)\phi\lo(\br')\ran+\lan\phi\lo(\br)\phi\s(\br')\ran.
\ee

In order to relate the 2PCFs in Eq.~(\ref{eq35}) to the pair distribution functions~(\ref{eq30}), we inject first into the SD identity~(\ref{eq24}) the Hamiltonian functional~(\ref{eq11}) and set $F=\phi_\gamma(\br')$. This yields
\bea
\label{eq36}
&&\hspace{-4mm}\int\hspace{-1mm}\mathrm{d}^3\br_1 v^{-1}_\gamma(\br,\br_1)\lan\phi_\gamma(\br')\phi_\gamma(\br_1)\ran\hspace{-.5mm}-\hspace{-.5mm}i\sum_{i=1}^p\lambda_iq_i\lan\hk_i(\br)\phi_\gamma(\br')\ran\nonumber\\
&&\hspace{-4mm}=\delta^3(\br-\br').
\eea
Inverting Eq.~(\ref{eq36}) via the identity
\be
\label{eq37}
\int\mathrm{d}^3\br_1v^{-1}_\gamma(\br,\br_1)v_\gamma(\br_1,\br')=\delta^3(\br-\br'),
\ee
the first term of Eq.~(\ref{eq35}) follows in the form
\bea\label{eq38}
\lan\phi_\gamma(\br)\phi_\gamma(\br')\ran&\hspace{-3mm}=\hspace{-3mm}&v_\gamma(\br,\br')\\
&&+i\sum_{i=1}^p\lambda_iq_i\hspace{-1mm}\int\hspace{-1mm}\mathrm{d}^3\br_1v_\gamma(\br,\br_1)\hspace{-.5mm}\lan\hk_i(\br_1)\phi_\gamma(\br')\ran.\nonumber
\eea
Then, in order to treat the cross terms in Eq.~(\ref{eq35}), we substitute again the Hamiltonian~(\ref{eq11}) into the SD Eq.~(\ref{eq24}), and set separately $F=\phi\lo(\br')$ for $\gamma={\rm s}$, and $F=\phi\s(\br')$ for $\gamma={\rm l}$. Using Eq.~(\ref{eq37}), this yields
\be\label{eq39}
\lan\phi_\gamma(\br)\phi_{\gamma'}(\br')\ran=i\sum_{i=1}^p\lambda_iq_i\hspace{-1mm}\int\hspace{-1mm}\mathrm{d}^3\br_1v_\gamma(\br,\br_1)\hspace{-.5mm}\lan\hk_i(\br_1)\phi_{\gamma'}(\br')\ran
\ee
for $\gamma={\rm s}$ and $\gamma'={\rm l}$, or $\gamma={\rm l}$ and $\gamma'={\rm s}$. Plugging now Eqs.~(\ref{eq38})-(\ref{eq39}) into Eq.~(\ref{eq35}), and accounting for the constraint~(\ref{eq6}), one obtains
\bea\label{eq40}
\mathcal G(\br,\br')&\hspace{-3mm}=\hspace{-3mm}&v\ce(\br,\br')\\
&&+i\hspace{-1mm}\sum_{\gamma=\{{\rm s,l}\}}\sum_{i=1}^p\lambda_iq_i\hspace{-1mm}\int\hspace{-1mm}\mathrm{d}^3\br_1v\ce(\br,\br_1)\hspace{-.5mm}\lan\hk_i(\br_1)\phi_\gamma(\br')\ran.\nonumber
\eea

Finally, in the SD identity~(\ref{eq24}), we set $F=\lambda_i\hk_i(\br_1)$ to obtain
\bea
\label{eq41}
&&\hspace{-1mm}\int\hspace{-1mm}\mathrm{d}^3\br_2 v^{-1}_\gamma(\br,\br_2)\lan\phi_\gamma(\br_2)\hk_i(\br_1)\ran-in_i\sum_{j=1}^pn_jq_iH_{ij}(\br,\br_1)\nonumber\\
&&\hspace{-1mm}=in_iq_i\delta^3(\br-\br_1),
\eea
where we used the global electroneutrality condition~(\ref{eq32}) and the definition of the total correlation function
\be
\label{eq42}
H_{ij}(\br,\br')=g_{ij}(\br,\br')-1.
\ee
The inversion of the identity~(\ref{eq41}) via Eq.~(\ref{eq37}) now yields
\be
\label{eq43}
\lambda_i\lan\hk_i(\br_1)\phi_\gamma(\br')\ran=in_i\int\mathrm{d}^3\br_2Q_i(\br_1,\br_2)v_\gamma(\br_2,\br'),
\ee
where we defined the net charge density
\be\label{eq44}
Q_i(\br,\br')=q_i\delta^3(\br-\br')+\sum_{j=1}^pn_jq_jH_{ij}(\br,\br')
\ee
of a central ion $q_i$ located at $\br$ dressed by its ionic atmosphere of radius $R=||\br-\br'||$. Plugging the identity~(\ref{eq43}) into Eq.~(\ref{eq40}), one finally obtains the following Ornstein-Zernike (OZ)-like identity 
\bea
\label{eq45}
\mathcal G(\br,\br')&\hspace{-3mm}=\hspace{-3mm}&v\ce(\br,\br')\\
&&-\hspace{-1mm}\sum_{i=1}^pn_iq_i\hspace{-1mm}\int\hspace{-1mm}\mathrm{d^3}\br_1\mathrm{d^3}\br_2v\ce(\br,\br_1)Q_i(\br_1,\br_2)v\ce(\br_2,\br')\nonumber
\eea 
relating the 2PCF and the total correlation function via the definition~(\ref{eq44}). We note in passing that in Appendix~\ref{pr}, via the use of the identities derived in this part, we report an alternative proof of the variational equation~(\ref{eq18}).

\subsection{Moment conditions}

We obtain here the electrostatic moment conditions required for the derivation of the screening parameter.

\subsubsection{Zeroth moment condition}

The zeroth moment condition~\cite{Hansen,AttardRev,Stir} complementing the global electroneutrality constraint~(\ref{eq32}) can be straightforwardly obtained by integrating Eq.~(\ref{eq41}) for  $\gamma={\rm l}$ over $\br$. Accounting for the cancellation of the inverse of the Fourier-transformed long-range potential~(\ref{eq15}) in the IR limit, i.e. $\tv\lo^{-1}(q\to0)=0$, this yields
\be
\label{eq46}
q_i+\sum_{j=1}^pn_jq_j\int\mathrm{d}^3\br H_{ij}(\br,\br\ce)=0.
\ee
Using now the definition of the local charge density~(\ref{eq44}), the zeroth moment constraint~(\ref{eq46}) can be equally expressed as the local electroneutrality condition around a central charge $q_i$ located at $\br\ce$, i.e.
\be
\label{eq47}
\int\mathrm{d}^3\br\;Q_i(\br,\br\ce)=0.
\ee

\subsubsection{Second moment condition}

Within the present formalism, we derive now the second moment condition originally obtained by Stillinger and Lovett via the exploitation of the electrical conductivity of the electrolyte~\cite{Stir}. To this aim, we account for the translational symmetry implying $\mathcal G(\br,\br')=\mathcal G(\br-\br')$, and carry out the FT of the 2PCF~(\ref{eq45}). This yields 
\be
\label{eq48}
\tgg(q)=\tv\ce(q)\left\{1-\tv\ce(q)\sum_{i=1}^pn_iq_i\tq_i(q)\right\},
\ee
where the FT of the charge density~(\ref{eq44}) reads
\be
\label{eq49}
\tq_i(q)=q_i+\sum_{j=1}^pn_jq_j\int\mathrm{d}^3\br H_{ij}(r)\frac{\sin(qr)}{qr}.
\ee

According to Eq.~(\ref{eq48}), the dielectric spectrum of the liquid defined as $\tgg(q)=:\tv\ce(q)/\tep(q)$ reads
\be
\label{eq50}
\tep(q)=\left\{1-\tv\ce(q)\sum_{i=1}^pn_iq_i\tq_i(q)\right\}^{-1}.
\ee
The electrical conductivity of the solution implying its perfect screening ability at large distances requires precisely the IR divergence of the spectrum~(\ref{eq50}), i.e. $\tep(q\to0)\to\infty$. In order to identify the corresponding constraint, we Taylor-expand the function~(\ref{eq50}) and account for the zeroth moment condition~(\ref{eq46}) to obtain
\be
\label{eq51}
\tep(q)=\left\{1+\frac{2\pi\ell_{\rm B}}{3}\sum_{i=1}^pn_iq_i\left[I_i^{(2)}-\frac{I_i^{(4)}}{20}q^2\right]+O(q^4)\right\}^{-1},
\ee
where we defined the moment integrals 
\be
\label{eq52}
I_i^{(n)}=\sum_{j=1}^pn_jq_j\int\mathrm{d}^3\br r^nH_{ij}(r).
\ee
Noting that the IR divergence of the spectrum necessitates the first two terms in Eq.~(\ref{eq51}) to cancel out, one finally obtains the second moment condition~\cite{Hansen,AttardRev,Stir}
\be\label{eq53}
\frac{2\pi\ell_{\rm B}}{3}\sum_{i,j}n_in_jq_iq_j\int\mathrm{d}^3\br \;r^2H_{ij}(r)=-1.
\ee

\subsection{Screening parameter}

\label{scr}

In order to calculate the screening parameter $\kappa$, we express first the inverse Fourier transform of Eq.~(\ref{eq48}),
\be
\label{eq54}
\mathcal G(\br-\br')=\int\frac{\mathrm{d}^3\bq}{(2\pi)^3}\frac{4\pi\ell_{\rm B}e^{i\bq\cdot(\br-\br')}}{q^2-4\pi\ell_{\rm B}\tchi(q)},
\ee
with the electric susceptibility defined by the identity $\tgg^{-1}(q)=:\tv\ce^{-1}(q)-\tchi(q)$, or
\be
\label{eq55}
\tchi(q)=\frac{\sum_{i=1}^pn_iq_i\tq_i(q)}{\tv\ce(q)\sum_{i=1}^pn_iq_i\tq_i(q)-1}.
\ee
Then, we assume the exponential decay of the correlation function at large distances, i.e.
\be\label{eq56}
\mathcal G(\br-\br')\approx\frac{\ell_{\rm B}e^{-\kappa ||\br-\br'||}}{\e^*||\br-\br'||}=\int\frac{\mathrm{d}^3\bq}{(2\pi)^3}\frac{4\pi\ell_{\rm B}e^{i\bq\cdot(\br-\br')}}{\e^*\left(q^2+\kappa^2\right)},
\ee
where the ionic correlation-dressed dielectric coefficient $\e^*$ renormalizes the pure solvent permittivity $\ew$~\cite{Kj2}.  

The comparison of Eqs.~(\ref{eq54}) and~(\ref{eq56}) indicates the presence of poles satisfying the characteristic equation 
\be
\label{eq57}
\kappa^2=-4\pi\ell_{\rm B}\tchi(i\kappa)
\ee
originally derived by Kjellander within the dressed-charge formalism~\cite{Kj1995}. In the present work, we will limit ourselves to the leading order perturbative solution of Eq.~(\ref{eq57}), i.e. $\kappa^2\approx-4\pi\ell_{\rm B}\tchi(0)$. Evaluating the IR limit of the Fourier-transformed susceptibility~(\ref{eq55}) together with the FT of the local net charge~(\ref{eq49}), and accounting for the moment conditions~(\ref{eq46}) and~(\ref{eq53}), one finally obtains 
\be
\label{eq58}
\kappa^2\approx-\frac{30}{\pi\ell_{\rm B}\sum_{i=1}^pn_iq_iI_i^{(4)}}. 
\ee

\subsection{Integral relation between the 2PCF and the pair distribution functions}

The direct evaluation of the 2PCF via the Fourier integral~(\ref{eq54}) is a numerically demanding task. Thus, we obtain here the 2PCF in terms of the total correlation functions~(\ref{eq42}) via a real space integral practical for numerical evaluation. To this aim, we use Eq.~(\ref{eq49}) together with the moment conditions~(\ref{eq46}) and~(\ref{eq53}) to obtain
\be
\label{eq59}
\sum_{i=1}^pn_iq_i\tq_i(q)=\tv\ce^{-1}(q)-\hspace{-.5mm}\int\hspace{-1mm}\mathrm{d}^3\br \hspace{.5mm}T(r)\hspace{-1mm}\left[\frac{\sin(qr)}{qr}+\frac{(qr)^2}{6}-1\right],
\ee
with the charge-charge correlation function defined as 
\be
\label{eq60}
T(r)=-\sum_{i,j}n_in_jq_iq_jH_{ij}(r).
\ee
Via Eq.~(\ref{eq59}), the FT of the 2PCF in Eq.~(\ref{eq48}) can be now reduced to
\be
\label{eq61}
\tgg(q)=\tv\ce^2(q)\int\mathrm{d}^3\br\;T(r)\left[\frac{\sin(qr)}{qr}+\frac{(qr)^2}{6}-1\right].
\ee
Next, we take the inverse FT of Eq.~(\ref{eq61}) to obtain
\be
\label{eq62}
\mathcal G(r)=2\pi\ell_{\rm B}^2\int\mathrm{d}^3u\;T(u)\left(r+\frac{u^2}{3r}-\delta u\right),
\ee
where $\delta u=\left(r^2+u^2-2ur\cos\theta_u\right)^{1/2}$. Evaluating in Eq.~(\ref{eq62}) the integral over the solid angle, one finally obtains a one-dimensional integral relation between the 2PCF and the total correlation functions,
\be
\label{eq63}
\mathcal G(r)=-\frac{8\pi^2\ell_{\rm B}^2}{3r}\sum_{i,j}n_in_jq_iq_j\int_r^\infty\mathrm{d}uu(u-r)^3H_{ij}(u).
\ee
It is noteworthy that via Eq.~(\ref{eq42}), the exact identity~(\ref{eq63}) can be used to compute the 2PCFs from the ionic pair distributions extracted from numerical simulations.

\subsection{Evaluation of the statistical averages}

\subsubsection{Mixed expansion scheme}
\label{exsc}

Due to the non-linearity of the functional Hamiltonian~(\ref{eq11}), the functional averages involved in the identities derived above cannot be evaluated exactly. Within the SCDH formalism, these statistical averages are calculated via the variationally augmented virial and cumulant treatment of the short- and long-range ion interactions, respectively. This mixed approximation scheme is based on the exact splitting of the Hamiltonian~(\ref{eq11}) in the form
\be
\label{eq64}
H=H_0+t\delta H.
\ee
In Eq.~(\ref{eq64}), the reference Hamiltonian
\bea
\label{eq65}
\hspace{-5mm}\beta H_0&=&\int\frac{\mathrm{d}^3\br\mathrm{d}^3\br'}{2}\left\{\sum_{\gamma=\{{\rm s,h}\}}\phi_\gamma(\br)v_\gamma^{-1}(\br,\br')\phi_\gamma(\br')\right.\\
&&\hspace{3cm}\left.+\phi\lo(\br)G\lo^{-1}(\br,\br')\phi\lo(\br')\right\}.\nonumber
\eea
to be treated exactly accounts for the variance of the potential fluctuations associated with the short-range interactions ($\gamma=\{{\rm h,s}\}$) with the bare pairwise kernels in Eqs.~(\ref{eq0}) and~(\ref{eq16}). Moreover, Eq.~(\ref{eq65}) incorporates the long-range interactions ($\gamma={\rm l}$) via the 2PCF~(\ref{eq19})  to be derived from the SC solution of the SD Eq.~(\ref{eq38}), i.e.
\bea\label{eq65II}
G\lo(\br,\br')&\hspace{-3mm}=\hspace{-3mm}&v\lo(\br,\br')\\
&&+i\sum_{i=1}^p\lambda_iq_i\hspace{-1mm}\int\hspace{-1mm}\mathrm{d}^3\br_1v\lo(\br,\br_1)\hspace{-.5mm}\lan\hk_i(\br_1)\phi\lo(\br')\ran.\nonumber
\eea

The splitting~(\ref{eq64}) includes the expansion parameter $t$ of unit magnitude ($t=1$) that will allow to keep track of the perturbative order. From Eqs.~(\ref{eq11}) and~(\ref{eq65}), the corresponding Hamiltonian component to be treated perturbatively follows in the form
\bea
\label{eq66}
\beta\delta H&=&\int\frac{\mathrm{d}^3\br\mathrm{d}^3\br'}{2}\phi\lo(\br)\left[v\lo^{-1}(\br,\br')-G^{-1}\lo(\br,\br')\right]\phi\lo(\br')\nonumber\\
&&-\sum_{i=1}^p\lambda_i\int\mathrm{d}^3\br\;\hk_i(\br).
\eea
Finally, substituting the splitting~(\ref{eq64}) into Eq.~(\ref{eq20}), and Taylor-expanding the result in terms of the expansion parameter $t$, the statistical average of the general functional $F[\ph]$ follows at the first order cumulant-level as
\be
\label{eq67}
\lan F\ran=\lan F\ran_0-t\left[\lan\beta\delta H F\ran_0-\lan\beta\delta H\ran_0\lan F\ran_0\right]+O\left(t^2\right),
\ee
where we defined the Gaussian-level functional average 
\be\label{eq68}
\lan F[\ph]\ran_0=\frac{1}{Z_0}\int\mathcal{D}\ph\;e^{-\beta H_0[\ph]}F[\ph]
\ee
including the partition function $Z_0=\int\mathcal{D}\ph\;e^{-\beta H_0[\ph]}$.

\subsubsection{Relating ion fugacities to concentrations}

In order to relate the ionic fugacities to the experimentally tunable salt concentrations, we carry out first the cumulant expansion~(\ref{eq67}) of Eq.~(\ref{eq25}) at the order $O(t)$,
\bea
\label{eq69}
\lan \hk_i(\br)\ran&=&\lan \hk_i(\br)\ran_0\\
&&-t\left[\lan\beta\delta H  \hk_i(\br)\ran_0-\lan\beta\delta H\ran_0\lan  \hk_i(\br)\ran_0\right].\nonumber
\eea
Evaluating now the gaussian averages in Eq.~(\ref{eq69}) according to Eq.~(\ref{eq68}), and inserting into the result the formal expansion of the long-ranged 2PCF,
\be
\label{eq70}
G\lo(\br,\br')=G_{{\rm l},0}(\br,\br')+tG_{{\rm l},1}(\br,\br')+O(t^2),
\ee
 after some algebra, one obtains at the order $O(t)$
\bea
\label{eq71}
n_i&=&\Lambda_i-t\Lambda_i\frac{q_i^2}{2} G_{{\rm l},1}(0)+t\Lambda_i\sum_{j=1}^p\Lambda_j\int\mathrm{d}^3\br\;h_{ij}(\br)\\
&&-tq_i^2\Lambda_i\int\frac{\mathrm{d}^3\br_1\mathrm{d}^3\br_2}{2}\left[G^{-1}_{{\rm l},0}(\br_1,\br_2)-v\lo^{-1}(\br_1,\br_2)\right]\nonumber\\
&&\hspace{3cm}\times G_{{\rm l},0}(\br,\br_1)G_{{\rm l},0}(\br,\br_2),\nonumber
\eea
where we defined the rescaled fugacity
\be
\label{eq72}
\Lambda_i=\lambda_i\;e^{-\frac{q_i^2}{2}\left[G_{{\rm l},0}(0)-v_{{\rm l},0}(0)\right]},
\ee
and the Mayer function
\be
\label{eq73}
h_{ij}(r)=e^{-v\h(r)-q_iq_j\left[G_{{\rm l},0}(r)+v\s(r)\right]}-1.
\ee
Finally, injecting into Eq.~(\ref{eq71}) the formal expansion $\Lambda_i=\Lambda_i^{(0)}+t\Lambda_i^{(1)}+O(t^2)$, and identifying the terms of different perturbative orders, at the order $O(t)$, the ionic fugacity follows in terms of the ion concentration as
\bea
\label{eq71II}
\Lambda_i&=&n_i+tn_i\frac{q_i^2}{2} G_{{\rm l},1}(0)-tn_i\sum_{j=1}^pn_j\int\mathrm{d}^3\br\;h_{ij}(\br)\\
&&+tq_i^2n_i\int\frac{\mathrm{d}^3\br_1\mathrm{d}^3\br_2}{2}\left[G^{-1}_{{\rm l},0}(\br_1,\br_2)-v\lo^{-1}(\br_1,\br_2)\right]\nonumber\\
&&\hspace{3cm}\times G_{{\rm l},0}(\br,\br_1)G_{{\rm l},0}(\br,\br_2).\nonumber
\eea

\subsubsection{Calculation of the long-range kernel}

We explain here the derivation of the long-range kernel via the SC solution of Eq.~(\ref{eq65II}). To this aim, we carry out the expansion~(\ref{eq67}) of the functional average in Eq.~(\ref{eq65II}), evaluate the resulting Gaussian averages according to Eq.~(\ref{eq68}), and replace the ion fugacities by salt concentrations via the identity~(\ref{eq71II}). Accounting for the expansion~(\ref{eq70}) of the Green's function, and expanding the result at the order $O(t)$, the components of the long-range kernel follow as
\bea
\label{eq71III}
G_{{\rm l},0}(\br-\br')&=&v\lo(\br-\br')\\
&&-\sum_{i=1}^pn_iq_i^2\int\hspace{-1mm}\mathrm{d}^3\br_1 v\lo(\br-\br_1)G_{{\rm l},0}(\br_1-\br');\nonumber\\
\label{eq71IV}
G_{{\rm l},1}(\br-\br')&=&-\sum_{i=1}^pn_iq_i^2\int\hspace{-1mm}\mathrm{d}^3\br_1 v\lo(\br-\br_1)G_{{\rm l},1}(\br_1-\br')\nonumber\\
&&-\sum_{i,j}n_in_jq_iq_j\int\hspace{-1mm}\mathrm{d}^3\br_1\mathrm{d}^3\br_2 v\lo(\br-\br_1)\\
&&\hspace{7mm}\times \left[h_{ij}(\br_1-\br_2)+q_iq_jG_{{\rm l},0}(\br_1-\br_2)\right]\nonumber\\
&&\hspace{7mm}\times G_{{\rm l},0}(\br_2-\br').\nonumber
\eea
In the derivation of Eq.~(\ref{eq71IV}), we used the inverse of the identity~(\ref{eq71III}) obtained via Eq.~(\ref{eq37}), i.e.
\be\label{eq71V}
G_{{\rm l},0}^{-1}(\br,\br')=v\lo^{-1}(\br,\br')+\sum_in_iq_i^2\delta(\br-\br').
\ee
Finally, Fourier-transforming Eqs.~(\ref{eq71III})-(\ref{eq71IV}), the components of the long-range kernel in Eq.~(\ref{eq70}) follow in reciprocal space as
\bea
\label{eq72II}
\tG_{{\rm l},0}(q)&=&\left\{\tv\lo^{-1}(q)+\sum_{i=1}^pn_iq_i^2\right\}^{-1};\\
\label{eq73II}
\tG_{{\rm l},1}(q)&=&-\tG_{{\rm l},0}^2(q)\sum_{i,j}n_in_jq_iq_j\left\{\may+q_iq_j\tG_{{\rm l},0}(q)\right\},\nonumber\\
\eea
where the FT of the Mayer function~(\ref{eq73}) reads
\bea
\label{eq74}
\may&\hspace{-2mm}=\hspace{-2mm}&-\frac{4\pi}{q^3}\left[\sin(qd)-qd\cos(qd)\right]\\
&&+4\pi\int_d^\infty\hspace{-2mm}\mathrm{d}rr^2\frac{\sin(qr)}{qr}\left\{e^{-q_iq_j\left[v\s(r)+G_{{\rm l},0}(r)\right]}-1\right\}.\nonumber
\eea

The knowledge of the component $G_{{\rm l},0}(r)$ in closed-form speeds up significantly the numerical computation of the Fourier-transformed Mayer function~(\ref{eq74}). For the derivation of this kernel in real space, we express first the inverse FT of Eq.~(\ref{eq72II}) together with Eq.~(\ref{eq15}) to obtain
\be
\label{eq78}
G_{{\rm l},0}(r)=\frac{2\ell_{\rm B}}{\pi r}\int_0^\infty\frac{\mathrm{d}qq\sin(qr)}{\kappa^2_{\rm DH}+q^2\left[1+(\sigma q)^2+(\sigma q)^4\right]},
\ee
with the DH screening parameter 
\be\label{dh}
\kappa^2_{\rm DH}=4\pi\ell_{\rm B}\sum_{i=1}^pn_iq_i^2.
\ee
Using the residue theorem, the Gaussian-level Green's function~(\ref{eq78}) can be expressed in closed-form as the sum 
\be
\label{eq79}
G_{{\rm l},0}(r)=\frac{\ell_{\rm B}}{r}\sum_{\gamma=\{0,\pm\}}\frac{e^{ik_ir/\sigma}}{3k_\gamma^4+2k_\gamma^2+1}
\ee
over the poles located in the upper half-plane,
\bea
\label{eq80}
k_0&=&\frac{1}{\sqrt3}\left\{\frac{u}{2^{1/3}}-\frac{2^{4/3}}{u}-1\right\}^{1/2};\\
k_\pm&=&\pm\frac{1}{\sqrt6}\left\{-2+\frac{2^{4/3}}{u}-\frac{u}{2^{1/3}}\right.\\
&&\hspace{1.2cm}\left.\pm i\left(\frac{2^{4/3}\sqrt{3}}{u}+\frac{\sqrt{3}u}{2^{1/3}}\right)\right\}^{1/2},\nonumber
\eea
where we introduced the auxiliary parameter $u=\left(7-27t^2+3^{3/2}\sqrt{27t^4-14t^2+3}\right)^{1/3}$ with $t=\kappa_{\rm DH}\sigma$.

\subsubsection{Evaluation of the variational equation~(\ref{eq18})} 

Within the mixed approximation scheme explained in Sec.~\ref{exsc}, we calculate now explicitly the variational equation~(\ref{eq18}) solved by the splitting length $\sigma$. To this aim, we evaluate first the 2PCF~(\ref{eq19}) for the short-range charge interactions via the SD Eq.~(\ref{eq38}) for $\gamma={\rm s}$, i.e.
\bea
\label{eq81}
G\s(\br,\br')&\hspace{-3mm}=\hspace{-3mm}&v\s(\br,\br')\\
&&+i\sum_{i=1}^p\lambda_iq_i\hspace{-1mm}\int\hspace{-1mm}\mathrm{d}^3\br_1v\s(\br,\br_1)\hspace{-.5mm}\lan\hk_i(\br_1)\phi\s(\br')\ran.\nonumber
\eea
In order to evaluate the functional average in Eq.~(\ref{eq81}), we carry out its cumulant expansion~(\ref{eq67}), calculate the Gaussian averages defined by Eq.~(\ref{eq68}), and express the ion fugacities in terms of the salt densities via Eq.~(\ref{eq71II}). Expanding the result at the order $O(t)$, one obtains
\bea
\label{eq82}
G\s(\br-\br')&=&v\s(\br-\br')\\
&&-\sum_{i=1}^pn_iq_i^2\int\hspace{-1mm}\mathrm{d}^3\br_1 v\s(\br-\br_1)v\s(\br_1-\br')\nonumber\\
&&-t\sum_{i,j}n_in_jq_iq_j\int\mathrm{d}^3\br_1\mathrm{d}^3\br_2v\s(\br-\br_1))\nonumber\\
&&\hspace{2.5cm}\times h_{ij}(\br_1-\br_2)v\s(\br_2-\br').\nonumber
\eea
Next, Fourier-expanding Eq.~(\ref{eq82}), one gets
\be
\label{eq83}
\tG\s(q)-\tv\s(q)\hspace{-.5mm}=\hspace{-.5mm}-\hspace{-1mm}\sum_{i=1}^pn_iq_i^2\tv\s^2(q)-t\hspace{-.5mm}\sum_{i,j}n_in_jq_iq_j\tv\s^2(q)\may.
\ee
Then, summing up Eqs.~(\ref{eq72II})-(\ref{eq73II}), the FT of the long-range 2PCF follows at the order $O(t)$ in the form
\bea
\label{eq84}
\tG\lo(q)-\tv\lo(q)&=&-\sum_{i=1}^pn_iq_i^2\tv\lo(q)\left[\tG_{{\rm l},0}(q)+t\tG_{{\rm l},1}(q)\right]\nonumber\\
&&-t\sum_{i,j}n_in_jq_iq_j\tv\lo(q)\tG_{{\rm l},0}(q)\nonumber\\
&&\hspace{8mm}\times\left[\may+q_iq_j\tG_{{\rm l},0}(q)\right].
\eea
Finally, inserting Eqs.~(\ref{eq83})-(\ref{eq84}) into the identity~(\ref{eq21}), after some algebra, the variational equation solved by the splitting parameter $\sigma$ follows at the order $O(t)$ as
\be\label{eq85}
\int_0^\infty\mathrm{d}qq^2\tG_{{\rm l},1}(q)\partial_\sigma\tv\lo(q)\left\{\tG_{{\rm l},0}^{-2}(q)-\tv\lo^{-2}(q)\right\}=0.
\ee
From Eq.~(\ref{eq85}), the value of the splitting parameter $\sigma$ can be easily obtained via a standard dichotomy algorithm.

\subsubsection{Calculation of the total correlation function}

Herein, we calculate the total correlation function~(\ref{eq42}) required for the computation of the thermodynamic functions. To this aim, we evaluate the cumulant expansion~(\ref{eq67}) of the pair distribution function~(\ref{eq30}), carry out the resulting Gaussian averages according to Eq.~(\ref{eq68}), account for the cumulant expansion~(\ref{eq70}) of the 2PCF, and replace the ion fugacities by salt concentrations via Eq.~(\ref{eq71II}). After lengthy algebra, one obtains
\be
\label{eq86}
H_{ij}(\br-\br')=h_{ij}(\br-\br')+t\left[h_{ij}(\br-\br')+1\right]T_{ij} (\br-\br'),
\ee 
where we introduced the auxiliary function 
\bea
 \label{eq87}
T_{ij}(\br-\br')&=&\sum_{k=1}^pn_k\int\mathrm{d}^3\br_1\left\{h_{ik}(\br-\br_1)h_{kj}(\br_1-\br')\right.\nonumber\\
&&\left.\hspace{1cm}-q_iq_jq_k^2G_{{\rm l},0}(\br-\br_1)G_{{\rm l},0}(\br_1-\br')\right\}\nonumber\\
&&-q_iq_jG_{{\rm l},1}(\br-\br').
\eea
Finally, passing to reciprocal space, the auxiliary function~(\ref{eq87}) can be expressed solely in terms of the Fourier-transformed functions~(\ref{eq72II})-(\ref{eq74}) to be obtained from the SC solution of the variational Eq.~(\ref{eq85}), i.e.
\bea
\label{eq88}
T_{ij}(r)&=&\int_0^\infty\frac{\mathrm{d}qq^2}{2\pi^2}\frac{\sin(qr)}{qr}\\
&&\hspace{7mm}\times\left\{\sum_{k=1}^pn_k\left[\tilde{h}_{ik}(q)\tilde{h}_{kj}(q)-q_iq_jq_k^2\tG_{{\rm l},0}^2(q)\right]\right.\nonumber\\
&&\left.\hspace{1.3cm}-q_iq_j\tG_{{\rm l},1}(q)\right\}.\nonumber
\eea

\subsection{Ion Association Model}
\label{ass}

In this article, ion association in symmetric solutions will be characterized within the framework of an upgraded version of the original ion pairing theory by Bjerrum~\cite{Bjerrum}. In this model, the association of the cation ${\rm C}^{q+} $ and the anion ${\rm A}^{q-} $ of valencies $\pm q$ into the neutral pair CA is formulated via the following equation of chemical equilibrium and mass action law~\cite{Marj,ExpSc},
\be
\label{eq91}
{\rm C}^{q+}+{\rm A}^{q-}\rightleftharpoons{\rm CA};\hspace{5mm}K=\frac{\left[{\rm AC}\right]}{\left[{\rm C}^{q+}\right]\left[{\rm A}^{q-}\right]}.
\ee
In Eq.~(\ref{eq91}), $K$ is the association constant, and $\left[{\rm C}^{q+}\right]=\left[{\rm A}^{q-}\right]=n_{{\rm ion}}$ and $\left[{\rm AC}\right]=n_{\rm p}$ are the concentrations of the dissociated ions and the neutral pairs, respectively. 

At this point, one takes into account the particle number conservation implying the constraint $n_{\rm p}=n_i-n_{ion}$. Combining the latter identity with the mass action law $K=n_{\rm p}/n_{ion}^2$ in Eq.~(\ref{eq91}), one obtains the characteristic equation $Kn_{ion}^2+n_{ion}-n_i=0$ whose solution yields the ionic pair fraction $\alpha=1-n_{\rm ion}/n_i$ in the form
\be\label{eq92}
\alpha=\frac{1+2u-\sqrt{1+4u}}{2u}.
\ee
In Eq.~(\ref{eq92}), we defined the dimensionless configurational integral $u=n_iK$ corresponding to the volume integral of the oppositely charged pair distribution function~\cite{Bjerrum,Marj},
\be\label{eq93}
u=4\pi n_i\int_d^\lambda\mathrm{d}rr^2g_{+-}(r),
\ee
where the cut-off length $\lambda$ is the interionic distance beyond which the electrostatically induced adhesion of the opposite charges $\pm q$ is ruptured by thermal energy. 

In the original ion pairing theory, the distance $\lambda$ is chosen as the boundary between the electrostatically dominated decaying branch ($r<\lambda$) and the entropy-driven rising branch ($r>\lambda$) of the radial distribution function $r^2g_{+-}(r)$ (see the inset of Fig.~\ref{fig9}(c))~\cite{Bjerrum}, i.e.
\be
\label{eq94}
\lambda=\min_{r}\left[r^2g_{+-}(r)\right].
\ee
As Bjerrum's formalism neglects the ionic atmosphere of the oppositely charged pair, it approximates the pair distribution function by the Boltzmann distribution associated with the bare Coulombic coupling of these ions, i.e. $g_{+-}(r)\approx e^{q^2\ell_{\rm B}/r}$. Within this approximation, the characteristic length~(\ref{eq94}) follows as $\lambda\approx\lambda_{\rm BJ}=q^2\ell_{\rm B}/2$. 

In the present work, the ionic pair fraction~(\ref{eq92}) will be computed via the calculation of the association rate~(\ref{eq93}) and the cut-off distance~(\ref{eq94}) with the pair distribution functions of the SCDH formalism provided by Eqs.~(\ref{eq42}) and~(\ref{eq86}) (see again the inset of Fig.~\ref{fig9}(c)). This improvement will allow to account for the ionic environment of the interacting pairs neglected by the Bjerrum theory.

\begin{figure}
\includegraphics[width=.9\linewidth]{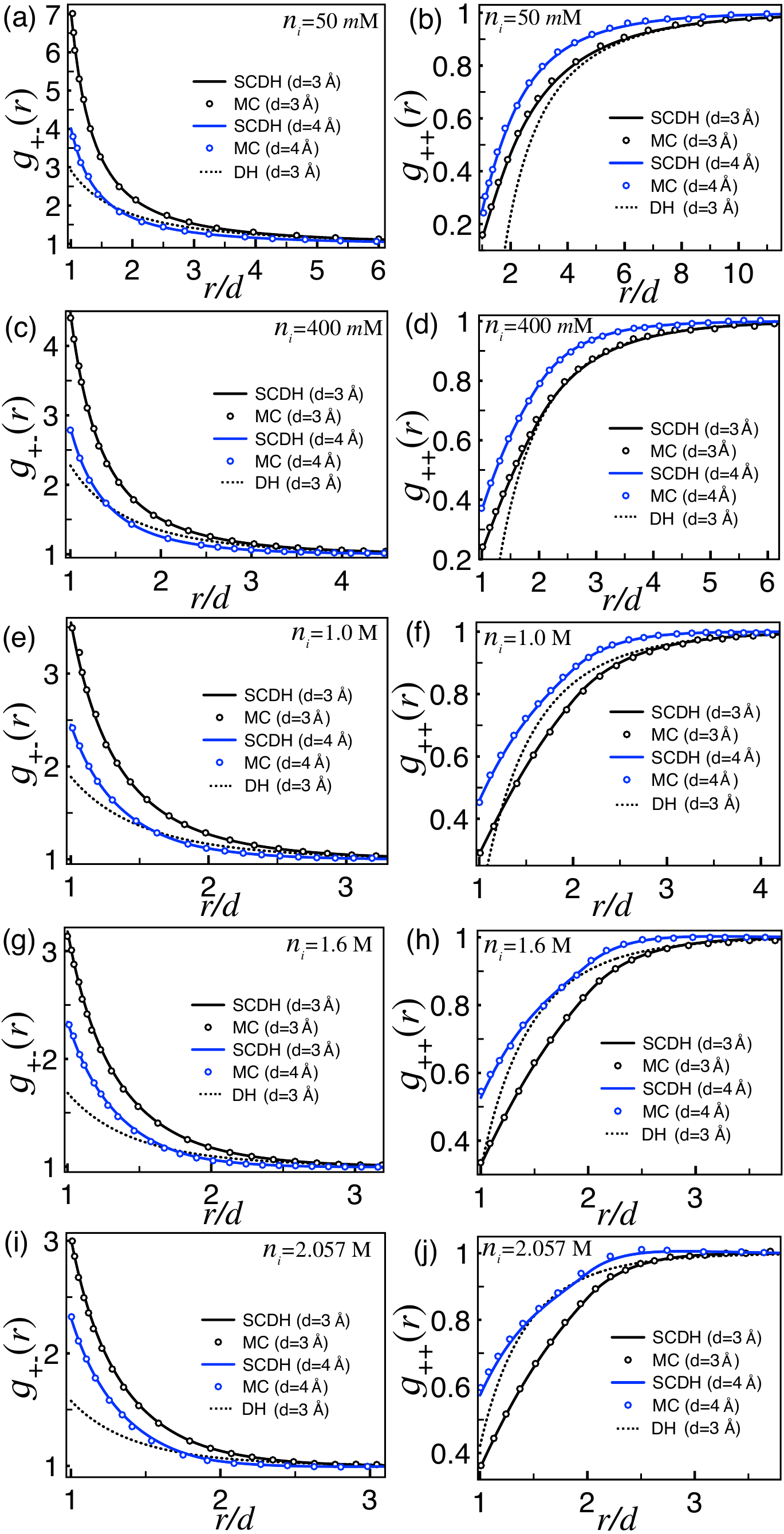}
\caption{(Color online) Cation-anion (left plots) and cation-cation (right plots) pair distributions at two HC sizes and various monovalent salt concentrations ($q_i=\pm1$). Solid curves: SCDH prediction from Eq.~(\ref{eq86}) via Eq.~(\ref{eq42}). Circles: MC data from Figs. 1 and 2 of the supplementary material of Ref.~\cite{Forsman}. Dotted curves: DH prediction $g_{ij}(r)=1-q_iq_j v_{\rm DH}(r)$ with $v_{\rm DH}(r)=\ell_{\rm B}e^{-\kappa_{\rm DH}r}/r$ for $d=3$ {\AA}~\cite{rem1}. The temperature and dielectric permittivity are $T=298$ K and $\ew=78.3$.}
\label{fig1}
\end{figure}

\section{Results}

\begin{figure}
\includegraphics[width=1.\linewidth]{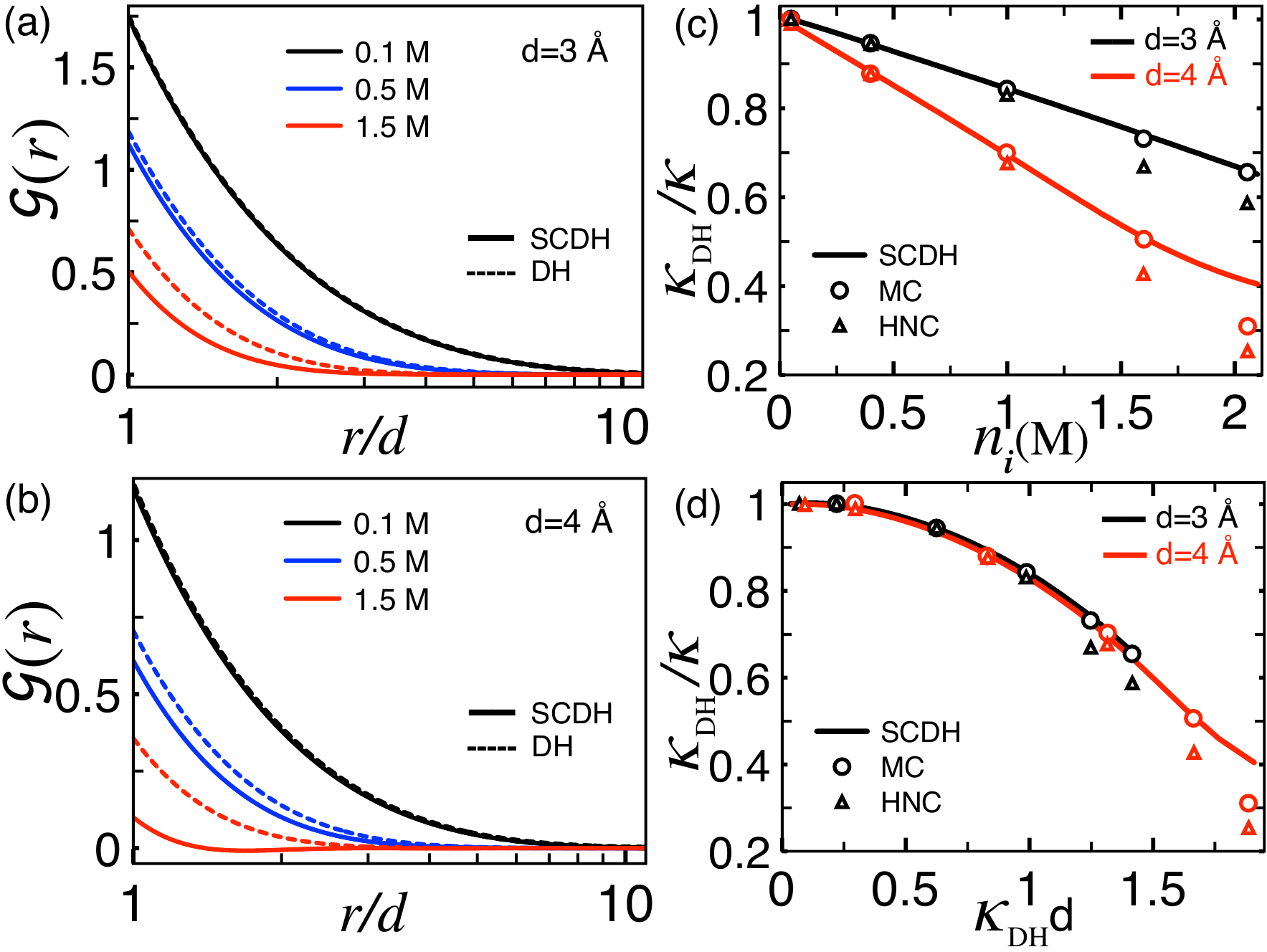}
\caption{(Color online) (a)-(b) Two-point correlation functions for the HC sizes $d=3$ {\AA} and $4$ {\AA}. Solid curves:  SCDH prediction~(\ref{eq63}). Dashed curves: DH potential $v_{\rm DH}(r)$. (c) Screening length rescaled by the DH length versus the salt concentration and (d) against the rescaled DH screening parameter. Solid curves: SCDH prediction~(\ref{eq58}). Symbols: MC and HNC data from Figs.11 (a)-(b) of Ref.~\cite{Forsman}. The temperature and dielectric permittivity of the monovalent electrolyte ($q_i=\pm1$) are $T=298$ K and $\ew=78.3$ in all figures.}
\label{fig2}
\end{figure}

In this part, we compare the ionic pair distribution profiles and the thermodynamic functions of the liquid obtained from the SCDH theory with numerical simulation results and experimental data from the literature.

\subsection{Comparison with MC simulations}

\label{compMC}

\subsubsection{Radial distribution functions}

Fig.~\ref{fig1} displays the opposite-charge (left plots) and like-charge (right plots) pair distributions of a monovalent electrolyte solution at the HC sizes $d=3$ {\AA} (black) and $4$ {\AA} (blue). One sees that from the dilute limit $n_i=50$ mM characterized by negligible HC interactions up to the molar regime $n_i=2.057$ M dominated by HC correlations, the SCDH predictions (solid curves) are in very good quantitative agreement with the MC data (circles) for both HC sizes. However, the DH prediction that neglects the ion size and embodies the assumption of linear electrostatic response is inaccurate in predicting the pair distributions over the whole density range including the dilute concentration regime~\cite{rem1}.

Figs.~\ref{fig2}(a)-(b) compare for the same HC sizes the 2PCFs of the SCDH formalism (solid curves) obtained from Eqs.~(\ref{eq63}) and~(\ref{eq86}) with the DH prediction (dashed curves). The plots show that  the DH approximation for the 2PCF can reproduce the SCHD result up to the concentration $n_i\sim0.5$ M. Thus, owing to the cancellation of errors upon summation over the ion species in Eq.~(\ref{eq63}), the DH theory can predict the 2PCFs in a more accurate fashion than the ionic pair distributions of Fig.~\ref{fig1}. The comparison of Figs.~\ref{fig2}(a) and (b) also shows that the inaccuracy of the DH theory amplifies with the HC size.

In Figs.~\ref{fig2}(c)-(d), we report the SCDH prediction~(\ref{eq58}) for the screening length (solid curves) corresponding to the range of the 2PCFs in (a)-(b) together with simulation data (circles) and HNC predictions (triangles). The comparison of the SCDH and MC results indicates that at the monovalent ion sizes $d=3$ {\AA} and $4$ {\AA}, the present theory can accurately reproduce the overscreening effect ($\kappa>\kappa_{\rm DH}$) originating from the finite interionic approach distance ($d\uparrow\kappa\uparrow$) up to the concentrations of $n_i\approx2.06$ M  and $1.6$ M, respectively~\cite{Kj2020,Soft2024}. One also notes that in the molar concentration regime, the SCDH prediction exhibits a better accuracy than the HNC approach.

\begin{figure}
\includegraphics[width=1.\linewidth]{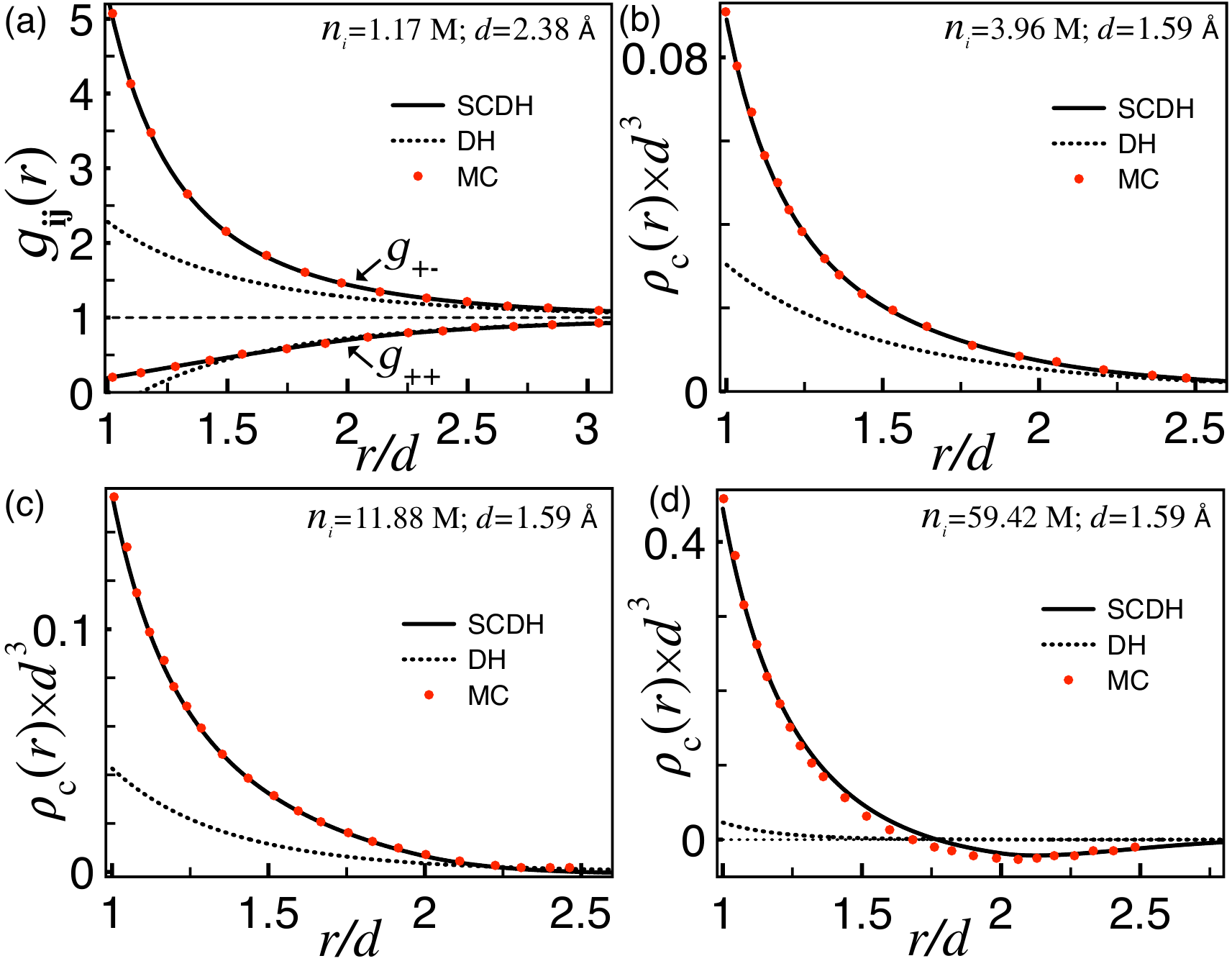}
\caption{(Color online) (a) Pair distribution functions, and (b)-(d) ionic charge density $\rho\ce(r)=q_+n_+H_{-+}(r)+q_-n_-H_{--}(r)$ around a central anion in 1:1 solutions. In (a), the reduced ion size is $d/\ell_{\rm B}=1/3$, and the packing fraction is $\eta=\pi n_id^3/3=0.01$. The MC data are from Fig.1 of Ref.~\cite{NetzMC}. In (b)-(d), the packing fractions are respectively $\eta=0.01$, $0.03$, and $0.15$. The reduced ion size is $d/\ell_{\rm B}=2/9$. The MC data are from Fig. 8B of Ref.~\cite{NetzMC}. The physical ion sizes and concentrations corresponding to the temperature $T=298$ K and dielectric permittivity $\ew=78.5$ are indicated in the legends.}
\label{fig3}
\end{figure}

Fig.~\ref{fig3} displays the pair distribution functions and the charge densities around a central anion at two considerably smaller ion sizes where the electrostatic interaction strength is substantially amplified by the closer interionic approach distances. The plots show that from the ionic packing fraction value $\eta=0.01$ located in the dilute salt regime into the concentrated regime $\eta=0.15$, the SCDH theory exhibits a good quantitative agreement with the MC simulation results. Even in the atypically small ion size and large concentration regime of Fig.~\ref{fig3}(d) where the tight competition between the electrostatic and HC correlations leads to local charge inversion, the SCDH formalism can reproduce the corresponding effect with reasonable precision. Finally, the DH theory is again inaccurate at all concentrations considered in the figure. 

\subsubsection{Thermodynamic functions}

We consider now the effect of ion size on the internal energy and pressure of the electrolyte. In Appendices~\ref{apen} and~\ref{appr}, we review the derivation of these functions solely in terms of the total correlation function~(\ref{eq86}) as
\bea
\label{eq89}
\beta E_{\rm ex}&=&2\pi\ell_{\rm B}\sum_{i=1}^p\sum_{j=1}^pn_in_jq_iq_j\int_d^\infty\mathrm{d}rrH_{ij}(r);\\
\label{eq90}
\beta P&=&\sum_{i=1}^pn_i+\frac{2\pi d^3}{3}\left(\sum_{i=1}^pn_i\right)^2\\
&&+\frac{2\pi d^3}{3}\sum_{i=1}^p\sum_{j=1}^pn_in_jH_{ij}(d^+)+\frac{1}{3}\beta E_{\rm ex}.\nonumber
\eea

The excess energy in Eq.~(\ref{eq89}) is composed of typical pairwise Coulomb interactions weighted by the correlation function~(\ref{eq86}). Then, on the r.h.s. of Eq.~(\ref{eq90}), the second term next to the ideal gas pressure is an excluded volume term corresponding to the first order virial expansion of the Carnahan–Starling (CS) pressure~\cite{Buyuk2024}. Moreover, the third term in Eq.~(\ref{eq90}) is the contribution of the contact pair densities including the second order virial expansion component of the CS pressure~\cite{Buyuk2024}. Finally, these repulsive pressure components are counterbalanced by the attractive energy contribution (fourth term) dominated by the Coulombic coupling of opposite charges.

\begin{figure}
\includegraphics[width=1.\linewidth]{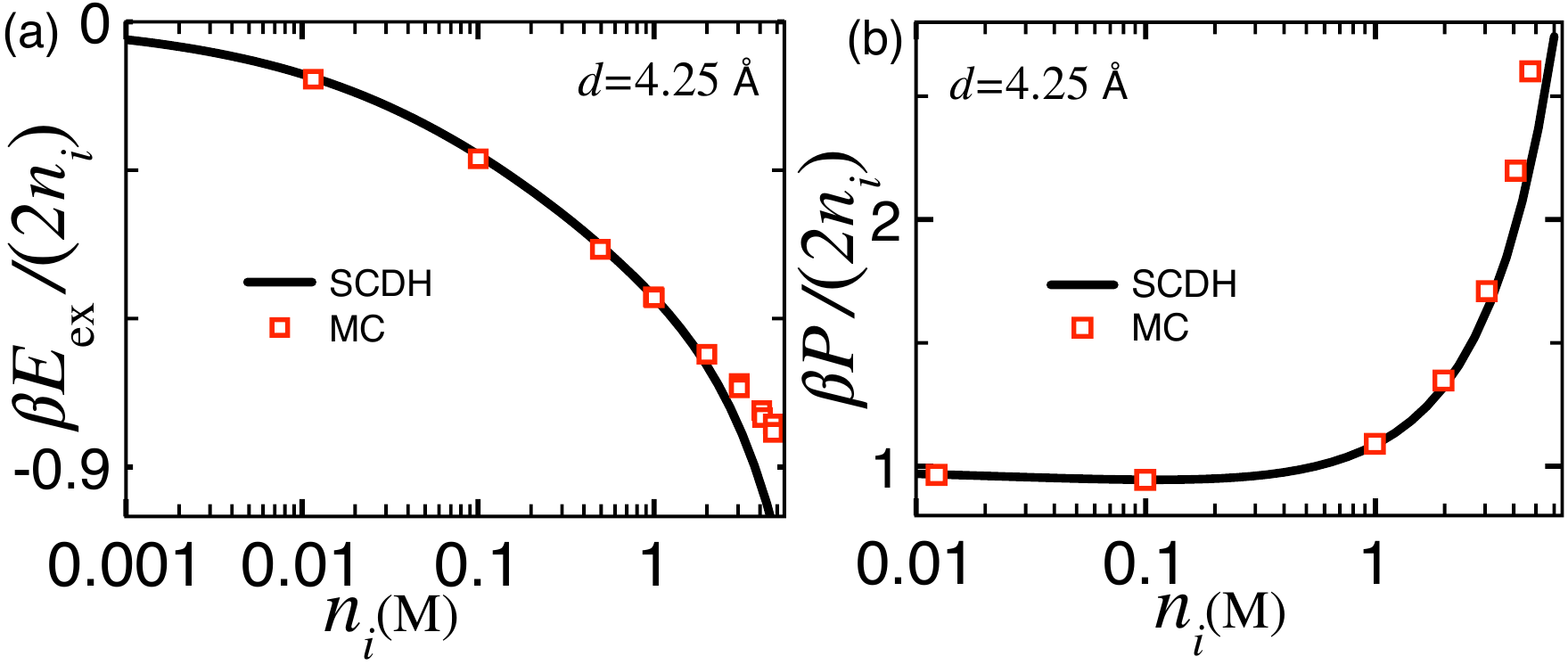}
\caption{(Color online) (a) Energy and (b) pressure against the monovalent ion concentration ($q_i=\pm1$) at the temperature $T=298$ K, the dielectric constant $\ew=78.5$, and the HC size $d=4.25$ {\AA}. Solid curves: SCDH predictions from Eqs.~(\ref{eq89})-(\ref{eq90}).  Symbols: MC data from (a) Ref.~\cite{Val80} and (b) Ref.~\cite{Svensson}.}
\label{fig4}
\end{figure}

Figs.~\ref{fig4}(a)-(b) display the excess energy and the osmotic coefficient of monovalent electrolytes against salt concentration at the HC size $d=4.25$ {\AA} close to the ionic size range of Figs.~\ref{fig1} and~\ref{fig2}. Consistent with the latter figures, one sees that the SCDH predictions exhibit good quantitative agreement with the MC data up to the molar concentration $n_i\approx2.0$ M. At larger concentrations where the SCDH theory overestimates the attractive energy and thus underestimates the pressure, our formalism can still reproduce the magnitude and trend of these thermodynamic functions with reasonable precision.

\begin{figure}
\includegraphics[width=1.\linewidth]{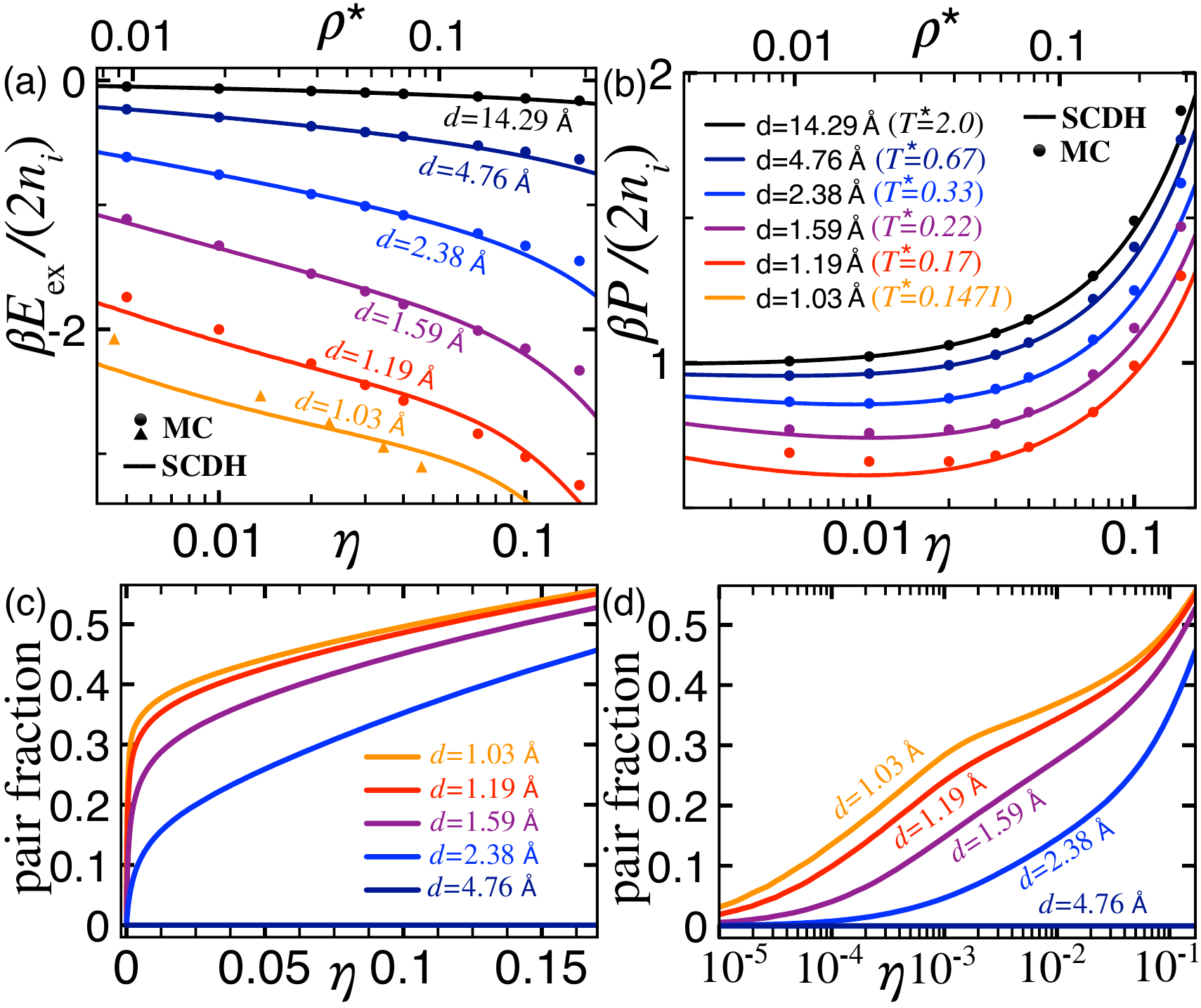}
\caption{(Color online) (a) Energy and (b) pressure against packing fraction $\eta=\pi n_id^3/3$ (lower horizontal axis) and reduced ion density $\rho^*=2n_id^3$ (upper horizontal axis) at various ion sizes. Solid curves: SCDH predictions from Eqs.~(\ref{eq89})-(\ref{eq90}). Disk symbols: MC data from (a) Table 1 and (b) Table 2 of Ref.~\cite{NetzMC}. Triangles in (a): MC data  from Fig.4 of Ref.~\cite{Val90}. From top to bottom, the reduced ion sizes or equivalently temperatures are $d/(\ell_{\rm B}q_i^2)=T^*=2$, $2/3$, $1/3$, $2/9$, $1/6$, and $0.1471$, respectively. The corresponding physical ion sizes for $T=298$ K, $\ew=78.5$, and $q_i=\pm1$ are given in the legends. (c) Pair fraction $\alpha$ from Eq.~(\ref{eq92}) on linear and (d) semi-logarithmic scales at the parameters of the upper plots.}
\label{fig5}
\end{figure}

In Figs.~(\ref{fig5})(a)-(b), we compare the SCDH predictions~(\ref{eq89})-(\ref{eq90}) for the internal energy and pressure with MC data at various HC sizes $d$ or equivalently reduced temperatures $T^*=d/\ell_{\rm B}$ (see the legend of (b)).  Therein, as the ion size is reduced from top to bottom, the weight of the attractive electrostatic interactions enhances with respect to the contribution of the repulsive HC interactions. This results in a more attractive internal energy ($d\downarrow E_{\rm ex}\downarrow$) and a less repulsive pressure ($d\downarrow P\downarrow$) equally developing an electrostatically dominated low density branch characterized by a weakly negative slope.

Figs.~(\ref{fig5})(a)-(b) show that within the packing fraction regime $n\lesssim0.1$, the SCDH theory can accurately account for this ion size dependence of the thermodynamic functions from $d=14.29$ {\AA} down to $d=1.59$ {\AA}. Then, the accuracy of the formalism deteriorates at the HC diameter $d=1.19$ {\AA}, and the approach looses its quantitative precision at the smaller ion size $d=1.03$ {\AA}. For monovalent solutions with HC size $d\approx3$ {\AA} and dielectric permittivity $\ew\approx 78.5$, the corresponding electrostatic coupling regime is reached at the excessively low solvent temperature of $T\approx 102$ K.

\begin{figure}
\includegraphics[width=1.\linewidth]{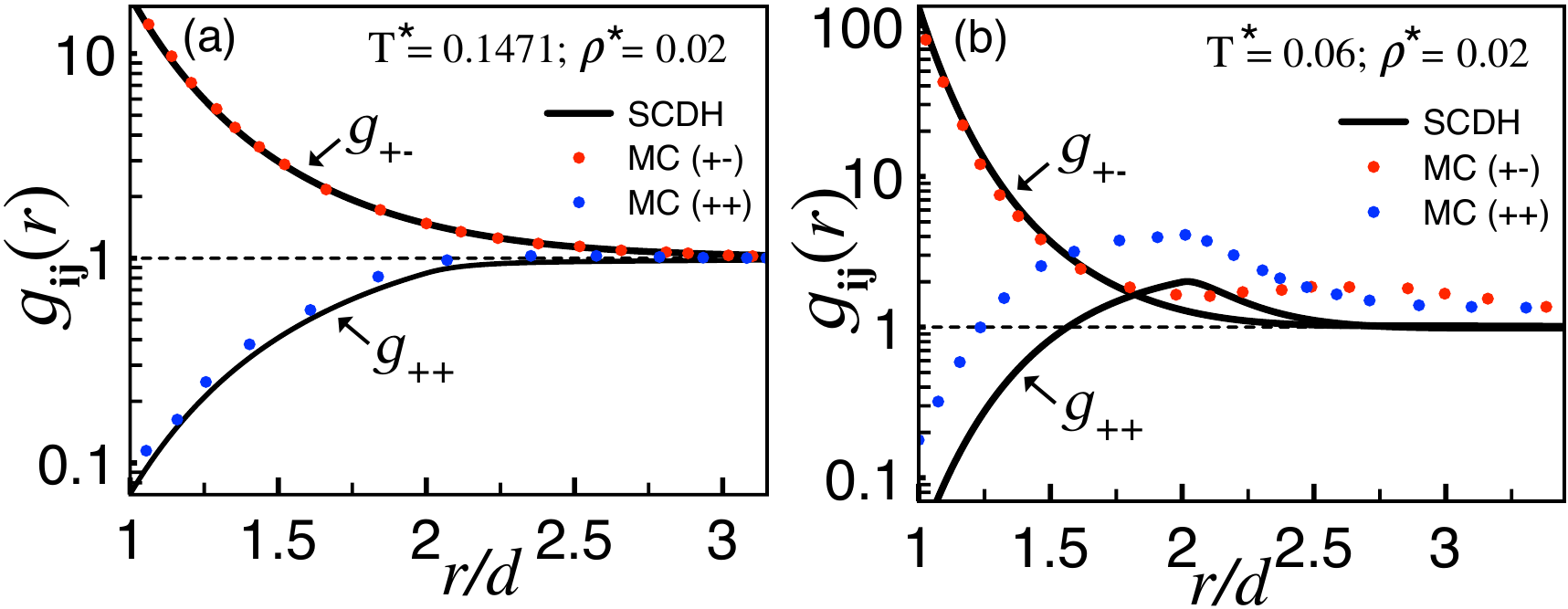}
\caption{(Color online) Cation-anion and cation-cation pair distribution functions at the reduced density $\rho^*=0.02$, and temperatures (a) $T^*=0.1471$ and (b) $T^*=0.06$. The MC data are from Figs. 3 and 4 of Ref.~\cite{Ork94}.}
\label{fig6}
\end{figure}

In order to shed light on the occurrence of this discrepancy at atypically small ion sizes or equivalently low temperatures, in Fig.~\ref{fig6}(a), we display the pair distribution functions at the corresponding electrostatic coupling strength. The comparison of the SCDH approach with MC data shows that the formalism can accurately reproduce the oppositely charged pair distribution but slightly underestimates the like-charge pair distribution. The underlying exaggeration of the like-charge interaction by the SCDH formalism is precisely responsible for the overestimation of the excess energy and the underestimation of the pressure in the small ion size and low concentration regime of Figs.~\ref{fig5}(a)-(b). Finally, Fig.~\ref{fig6}(b)  shows that at the considerably lower temperature $T^*=0.06$ comparable with the critical temperature of the L-V phase transition~\cite{Levin96},  the departure of the SCDH prediction from the MC data is substantially amplified. Thus, approaching the L-V critical point via the SCDH formalism will require the extension of the calculation scheme explained in Sec.~\ref{exsc} at least up to the next cumulant order.

\subsection{Comparison with experiments}
\label{expe}

In this part, we confront the osmotic coefficients calculated within the framework of the SCDH formalism with experimental data from the literature obtained for aqueous solutions and weakly polar non-aqueous electrolytes.

\subsubsection{Aqueous solutions}

\begin{figure}
\includegraphics[width=.97\linewidth]{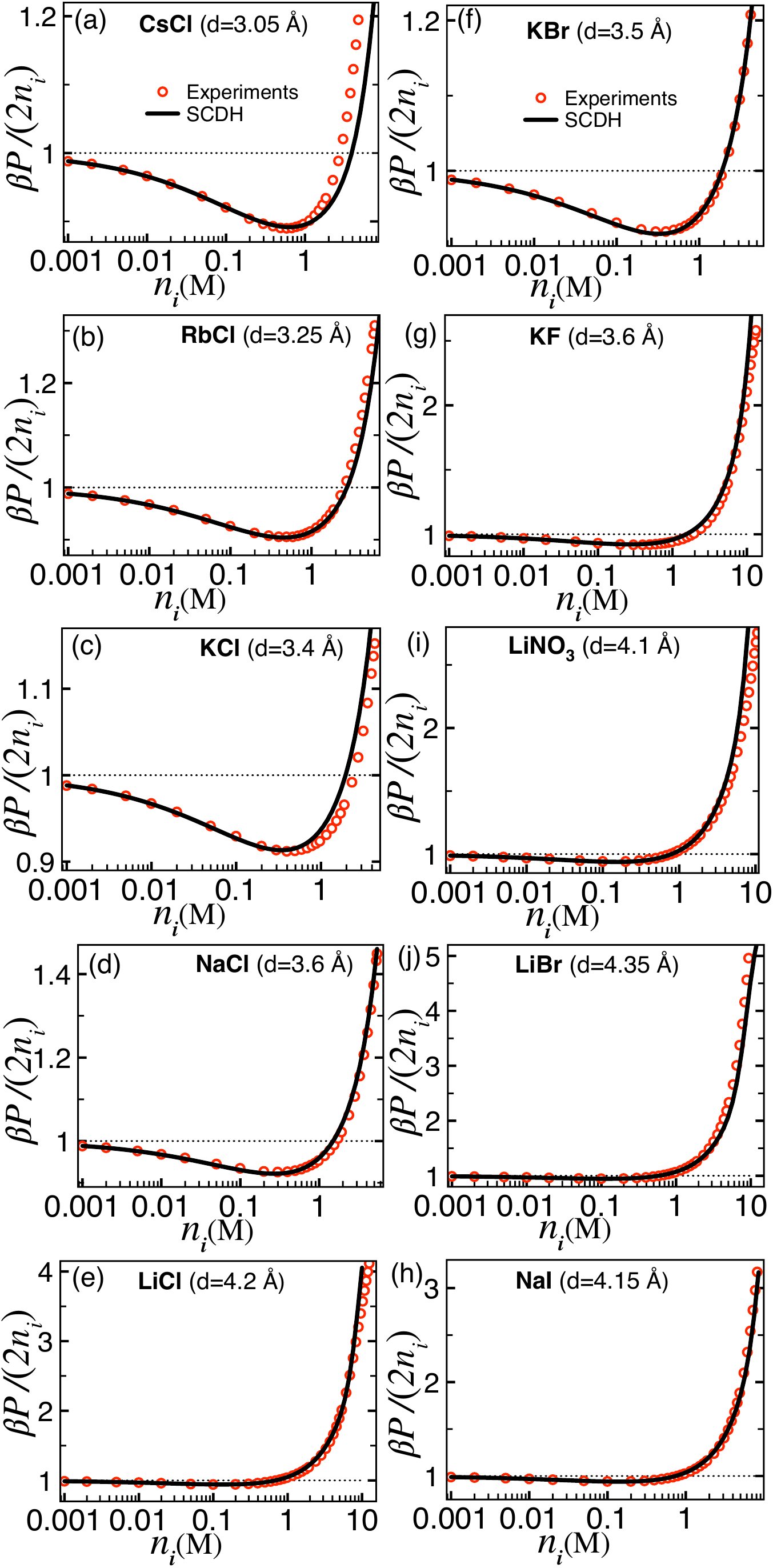}
\caption{(Color online) Osmotic coefficient of aqueous monovalent solutions at the temperature $T=298$ K. Solid curves: SCDH result~(\ref{eq90}). Circles: experimental data~\cite{rem2} from Ref.~\cite{Exp1}. The legends display the monovalent salt species and the corresponding HC size fitted to have the best agreement with the submolar branch of the experimental data. The salt-dependent dielectric permittivities are incorporated via the GP model~\cite{Gavish} explained in Appendix~\ref{Gav}.}
\label{fig7}
\end{figure}

In Fig.~\ref{fig7}, we reported the experimental osmotic coefficient of ten different aqueous electrolytes taken from Ref.~\cite{Exp1} (circles) together with SCDH results from Eq.~(\ref{eq90}) (solid curves). The conversion from the molal basis of the experiments to molar basis is explained in Appendix~\ref{conv}. In order to account for the substantial effect of salt-induced dielectric decrement, the osmotic coefficients have been computed by replacing the pure solvent permittivity $\ew$ of the SCDH theory with the salt-dependent permittivity $\e_{\rm el}$ of the Gavish-Promislow model (see Appendix~\ref{Gav})~\cite{Gavish}. Finally, the HC sizes $d$ have been adjusted to obtain the best agreement with the submolar branch of the experimental data. 

Fig.~\ref{fig7} shows that the experimental osmotic coefficients exhibit a non-uniform variation upon salt addition. This feature can be consistently characterized via the pressure identity~(\ref{eq90}). Namely, at submolar concentrations dominated by the effect of opposite charge attraction embodied by the forth term of Eq.~(\ref{eq90}), the amplification of this negative pressure component by salt increment lowers the osmotic coefficient, i.e. $n_i\uparrow\phi\downarrow$. Then, in the molar concentration regime where the ionic HC repulsion incorporated by the second and third terms of Eq.~(\ref{eq90}) take over electrostatic attraction, the osmotic coefficient rises with the salt concentration, i.e. $n_i\uparrow\phi\uparrow$. One sees that the SCDH formalism can accurately account for this non-monotonic salt dependence of the osmotic coefficient from submolar into molar concentrations. We also emphasize that the adjusted HC sizes of the chloride-based solutions in Figs.~\ref{fig7}(a)-(e) obey the cationic branch of the Hofmeister series, i.e.  ${\rm Cs}^+<{\rm Rb}^+<{\rm K}^+<{\rm Na}^+<{\rm Li}^+$~\cite{Hof1,Hof2}. 
\begin{figure*}
\includegraphics[width=1.\linewidth]{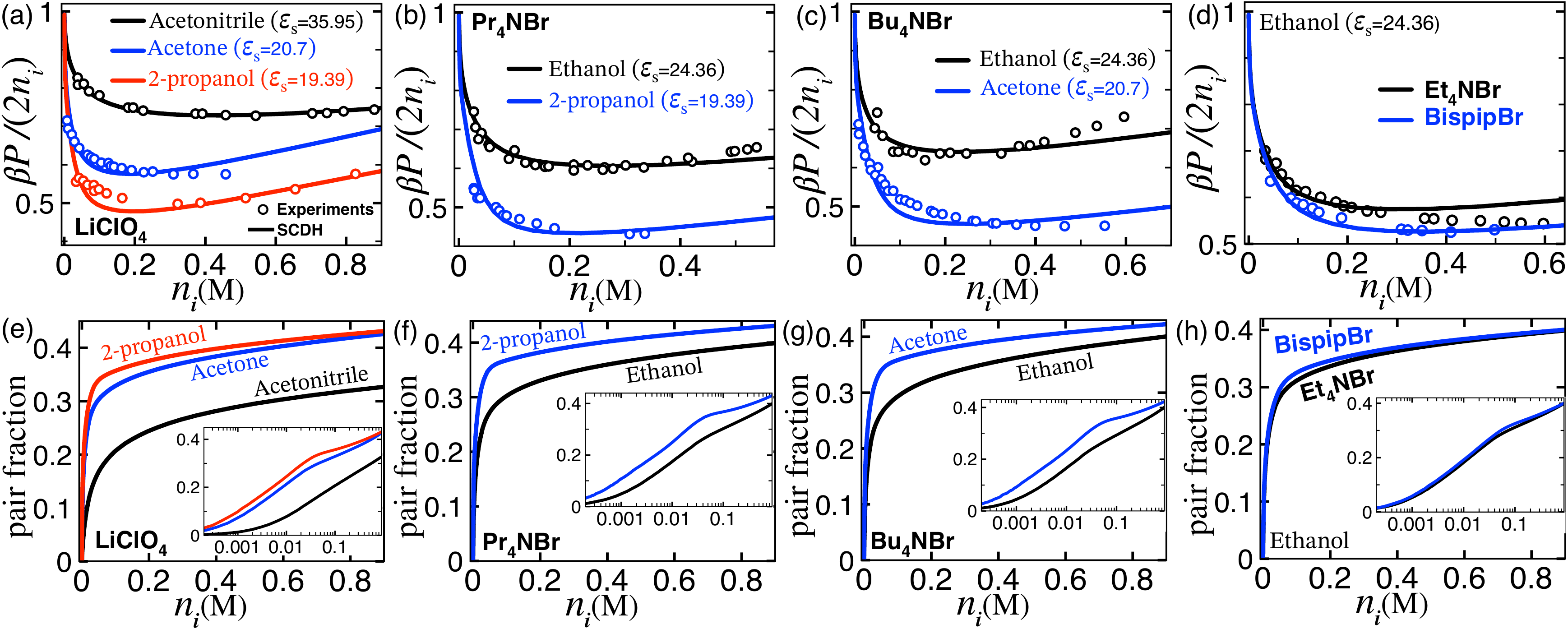}
\caption{(Color online) (a)-(d) Osmotic coefficient of non-aqueous monovalent solutions against solute concentration at the temperature $T=298$ K. Solid curves: SCDH prediction~(\ref{eq90}). Circles: experimental data from Ref.~\cite{Exp2}. The solvent and salt species, and the dielectric constant of each solvent taken from Ref.~\cite{Exp2} are indicated in the legends. (e)-(h) Pair fractions from Eq.~(\ref{eq92}) at the parameters of the upper plots. The insets display the main plots on a semi-logarithmic scale. The adjusted HC sizes are as follows: (a) $d=3.35$ {\AA} (acetonitrile), $d=4.25$ {\AA} (acetone), and $d=3.95$ {\AA} (2-propanol) for ${\rm LiClO}_4$; (b) $d=3.8$ {\AA} (ethanol) and $d=3.75$ {\AA} (2-propanol) for ${\rm Pr}_4{\rm NBr}$; (c) $d=4.05$ {\AA} (ethanol) and $d=3.6$ {\AA} (acetone) for ${\rm Bu}_4{\rm NBr}$; (d) $d=3.6$ {\AA} for ${\rm Et}_4{\rm NBr}$, and $d=3.35$ {\AA} for BispipBr in ethanol.}
\label{fig8}
\end{figure*}

\subsubsection{Non-aqueous solutions}

We consider now the case of non-aqueous electrolytes. In these fluids, the considerably low polarity of the solvent replacing water leads to the amplification of electrostatic correlations with respect to the aqueous solutions considered in Fig.~\ref{fig7}. In Figs.~\ref{fig8}(a)-(d), we reported the experimental osmotic coefficient of nine solutions characterized by low dielectric constants (circles). The experimental data readily transformed from molal to molar basis, and the corresponding dielectric permittivity values indicated in the legends have been taken from Ref.~\cite{Exp2}.

Figs.~\ref{fig8}(a)-(d) show that for a given type of solute, the osmotic coefficient decreases with the polarity of the solvent, i.e. $\ew\downarrow\phi\downarrow$.  According to the present formalism, this effect is induced by the enhancement of the forth term in the pressure identity~(\ref{eq90}); upon the reduction of the dielectric constant, the strengthening of opposite charge attraction amplifies this electrostatic pressure component of negative sign and drops the total pressure. The plots show that the SCDH theory (solid curves) can reproduce with reasonable accuracy  the corresponding decrease of the osmotic coefficient and the alteration of its non-uniform salt dependence with the variation of the dielectric permittivity~\cite{rem3}. In the next part, we probe the mapping between these thermodynamic features and the ion association phenomenon.

\subsection{Ion Association}
\label{ass}

In AFM force measurements conducted with non-aqueous charged solutions, the range of macromolecular interactions has been observed to be significantly longer than the DH screening length~\cite{ExpSc,ExpSc2,ExpAlc1}.  Within the framework of the ion association model introduced in Sec.~\ref{ass}, we characterize here the molecular mechanism of ionic pair formation at the basis of this underscreening effect.

\subsubsection{Effect of salt concentration}

Figs.~\ref{fig8}(e)-(h) display the ionic pair fraction~(\ref{eq92}) of the non-aqueous solutions in the upper panels. One sees that the monotonic rise of the pair fractions upon salt increment ($n_i\uparrow\alpha\uparrow$) occurs in two distinct regimes. In dilute solutions mainly governed by opposite charge attraction, the number of ion pairs sharply rise with the salt concentration up to the vicinity of the osmotic coefficient minima. Then, beyond the concentration regime $n_i\sim50$ mM where the excluded volume constraint becomes substantial, the underlying HC interactions hinder ion association and drive the pair fraction curves into a quasi-saturation state. 

Hence, due to the sharp competition between HC repulsion and the particularly strong opposite charge attraction set by the low solvent polarity, in non-aqueous solutions, substantial ion association occurs predominantly in the submolar salt range $n_i\lesssim50$ mM. This result is qualitatively consistent with AFM measurements indicating the rapid drop of ionization fractions in the corresponding millimolar concentration regime of non-aqueous electrolytes (see e.g. Fig. 5(a) of Ref.~\cite{ExpSc2}).

\subsubsection{Impact of solvent polarity and solute size}

\begin{figure}
\includegraphics[width=1.\linewidth]{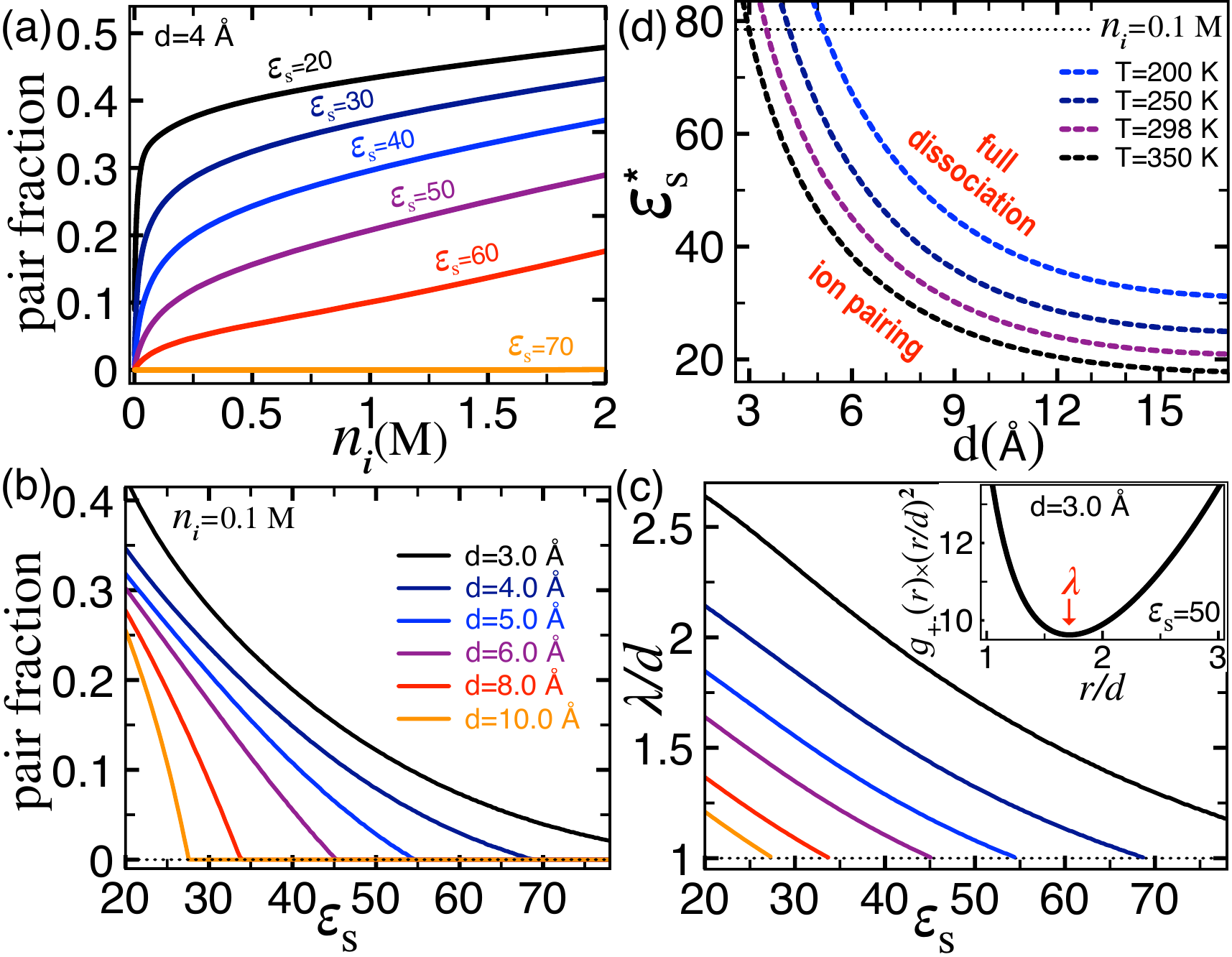}
\caption{(Color online) (a) Monovalent pair fractions ($q_i=\pm1$) from Eq.~(\ref{eq92}) against salt concentration at various dielectric constants and (b) versus dielectric permittivity at different HC sizes. (c) Maximum distance of opposite charge adhesion~(\ref{eq94}) rescaled by the HC size corresponding to the curve with the same color in (b). The inset displays the radial distribution function on the r.h.s. of Eq.~(\ref{eq94}). (d) Critical values of the dielectric constant where the pair fractions in (b) vanish against ion size at various temperatures.}
\label{fig9}
\end{figure}

In AFM experiments, the strength of ion association can be tuned via the polarity of the liquid by mixing alcohols with water solvent~\cite{ExpAlc1,ExpAlc2}. In particular, these experiments revealed that ion pairing is enhanced with the alcohol fraction of the solvent mixture~\cite{ExpAlc1}. In Fig.~\ref{fig9}(a), the corresponding effect is illustrated via the pair fraction curves computed at various dielectric permittivities. Indeed, one sees that the reduction of the dielectric constant amplifying electrostatic correlations strengthens ion association, i.e. $\e\s\downarrow\alpha\uparrow$. 

Fig.~\ref{fig9}(a) also shows that upon the increase of the dielectric constant to the value $\e\s=70$, the liquid becomes fully dissociated, i.e. $\alpha=0$. In Figs.~\ref{fig5}(c)-(d), this transition is displayed for aqueous electrolytes ($\e\s=78.5$) via the variation of the HC size. One sees that larger ion sizes giving rise to attenuated Coulomb coupling lead to weaker association, i.e. $d\uparrow\alpha\downarrow$. As a continuation of this trend, at the typical monovalent ion size $d=4.76$ {\AA}, ion pairing disappears entirely ($\alpha=0$). This is in line with AFM experiments indicating the absence of ion pairing in monovalent aqueous liquids considered in Fig.~\ref{fig7}~\cite{ExpSc,ExpSc2}. 

In Fig.~\ref{fig9}(b), the transition to the fully dissociated state of the fluid is illustrated in terms of the pair fractions against the solvent permittivity at various HC sizes. One notes that for each ionic size, the fraction of associated ions decays steadily with increasing dielectric constant and vanishes entirely beyond a characteristic permittivity value $\e\s^*$. With the aim to shed light on this feature, in Fig.~\ref{fig9}(c), we reported the maximum distance of pair adhesion~(\ref{eq94}) rescaled by the HC size. The plot shows that upon the increase of the dielectric constant towards the value $\e\s=\e\s^*$, the weakening of charge attraction reduces this adhesion radius ($\e\s\uparrow\lambda\downarrow$) down to the HC radius corresponding to the forbidden approach distance. Consequently, as the limit $\lambda=d$ is reached, the configurational integral~(\ref{eq93}) and the pair fraction~(\ref{eq92}) vanish. 

Hence, the condition $\lambda>d$ involving the upper pair distance~(\ref{eq94}) provides a quantitative criterion for the onset of ion association. In Fig.~\ref{fig9}(d), we plotted the isothermal lines $\e^*\s(d)$ set by the constraint  $\lambda=d$ separating the full dissociation phase from the ion pairing phase. The structure of this diagram is again dictated by the balance between charge attraction and core repulsion; the larger the ion size, the lower the maximum dielectric constant where ion pairing can be expected, i.e. $d\uparrow\e\s^*\downarrow$. As a result, at a given ion size $d$, the resurgence of ion association from the full dissociation phase requires the charge correlations suppressed by the HC repulsion to be recovered via the decrease of the temperature, i.e. $T\downarrow\e\s^*\uparrow$. 

\section{Conclusions}

Within the framework of an explicit ion size-augmented electrostatic formalism incorporating HC and charge interactions on an equal footing, we formulated a unified microscopic theory of equilibrium thermodynamics and ion association in charged solutions. Via a variational spitting technique enabling the asymmetric treatment of distinct interaction scales, our approach allows to avoid the WC treatment of strongly coupled short-range electrostatic and HC interactions. 

Comparison with recent simulation data~\cite{Forsman} showed that the SCDH theory can accurately reproduce the pair distributions of ions with the typical HC sizes $d=3$ {\AA} and $4$ {\AA} over the entire concentration range $50 \;{\rm  mM}\leq n_i\leq2.057$ M. Moreover, at the significantly smaller solute size $d=1.59$ {\AA} giving rise to strong electrostatic coupling, our formalism was shown to exhibit good quantitative agreement with the pair distributions of simulations~\cite{NetzMC} from the dilute regime $\eta=0.01$ up to the high packing regime $\eta=0.15$ governed by the tight competition between electrostatic and HC correlations. 

We identified as well the validity domain of our theory in terms of the ionic HC size. Systematic confrontation with simulations~\cite{NetzMC,Val90} showed that within the biologically relevant packing fraction regime $\eta\lesssim0.1$, the SCDH theory can accurately account for the solute size dependence of thermodynamic functions from $d\simeq14.3$ {\AA} down to $d\simeq1.6$ {\AA}. At smaller ion sizes equivalent to atypically low temperatures $T\lesssim102$ K, the overestimation of like-charge coupling by our cumulant approximation scheme leads to the break down of the SCDH formalism. In future works, the access to the corresponding regime of particularly strong electrostatic correlations covering the ionic L-V phase transition~\cite{Levin96} may be enabled by the extension of our calculation to higher cumulant orders. 

The predictions of our formalism have been compared with the experimental osmotic coefficient data of various aqueous and non-aqueous solutions. Via the only adjustment of the hydrated ion size, the SCDH theory can accurately account for the non-monotonic salt dependence of the experimental osmotic coefficients as well as its variation with the dielectric permittivity of the solvent from the submolar up to the molar concentration regime.

We also elucidated the molecular details of ion association responsible for the underscreening of macromolecular interactions in non-aqueous solutions~\cite{ExpSc,ExpSc2,ExpAlc1}. In qualitative agreement with AFM force measurements~\cite{ExpSc2}, we found that as a result of the particularly strong opposite charge attraction set by the low polarity of these fluids, substantial ion pairing occurs predominantly in the submolar concentration regime $n_i\lesssim50$ mM covering the sharply decreasing phase of the osmotic coefficient below the ideal gas limit; in the subsequent density regime $n_i\gtrsim50$ mM roughly coinciding with the rising phase of the osmotic coefficient, the underlying HC interactions hinder ion association and result in the quasi-saturation of the pair fraction curves upon salt increment.

The diagram in Fig.~\ref{fig9}(d) recapitulates our main conclusions on the impact of solvent polarity and hydration size on ionic pair formation. Namely, in aqueous solutions (dotted horizontal line), the occurrence of monovalent ion association at typical hydration sizes $d\gtrsim4$ {\AA} requires the lowering of the temperature below the freezing point of water. However, in the dielectric constant range $19\lesssim\e\s\lesssim 36$ and the temperature  $T=298$ K of the weakly polar liquids in Fig.~\ref{fig8}, the corresponding hydration radii coincide with the ionic association phase of the diagram. This picture is qualitatively consistent with the experimental observation of monovalent solute pairing exclusively in non-aqueous solvents and liquid mixtures with substantial alcohol content~\cite{ExpSc,ExpSc2,ExpAlc1}.

\smallskip
\appendix

\section{Derivation of the internal energy density and osmotic pressure}

In this appendix, we review the derivation of the excess energy and pressure of a general electrolyte mixture in terms of the total correlation functions~\cite{Hansen,Buyuk2024}. 

\subsection{Internal energy density}
\label{apen}

The excess energy density $E_{\rm ex}$ is defined as the GC averaged sum of the pairwise energy components~(\ref{eq2}) renormalized by the ionic self energy, i.e. 
\be\label{a0}
E_{\rm ex}=\frac{1}{V}\lan\left(E\ce+E\h\right)\ran_{\rm G}-\sum_{i=1}^pN_i\epsilon_i,
\ee
where $V$ is the total volume. Substituting into Eq.~(\ref{a0}) the identities~(\ref{eq2}) and~(\ref{eq4})-(\ref{eq5}), one obtains
\bea
\label{a1}
\beta E_{\rm ex}&=&\frac{1}{2V}\int\mathrm{d}^3\br\mathrm{d}^3\br'\sum_{i=1}^p\sum_{j=1}^p\left\{q_iq_jv\ce(\br,\br')+v\h(\br,\br')\right\}\nonumber\\
&&\times\lan\sum_{k=1}^{N_i}\sum_{l=1}^{N_j}\delta^3(\br-\br_{ik})\delta^3(\br-\br_{jl})\left(1-\delta_{ij}\delta_{kl}\right)\ran_{\rm G}.\nonumber\\
\eea
Comparing now the GC average term in Eq.~(\ref{a1}) with Eq.~(\ref{eq26}), one can express the excess energy in terms of the pair distribution function as
\bea
\label{a2}
\beta E_{\rm ex}&=&\int\frac{\mathrm{d}^3\br\mathrm{d}^3\br'}{2V}\sum_{i=1}^p\sum_{j=1}^pn_in_j\left[q_iq_jv\ce(\br,\br')+v\h(\br,\br')\right]\nonumber\\
&&\hspace{2.7cm}\times g_{ij}(\br,\br').
\eea

At this point, accounting for the translational symmetry in the system together with the property $g_{ij}(r)\propto\theta(r-d)$ originating from the HC constraint, Eq.~(\ref{a2}) reduces to
\be
\label{a3}
\beta E_{\rm ex}=2\pi\sum_{i=1}^p\sum_{j=1}^pn_in_j\int_d^\infty\mathrm{d}rr^2\left[q_iq_jv\ce(r)+v\h(r)\right]g_{ij}(r).
\ee
According to the definition~(\ref{eq0}), the HC potential vanishes outside the contact sphere, i.e. $v\h(r>d)=0$. This implies that the contribution from the HC potential to the integral in Eq.~(\ref{a3}) vanishes as well. This finally yields the internal energy density in the form
\be
\label{a40}
\beta E_{\rm ex}=2\pi\sum_{i=1}^p\sum_{j=1}^pn_in_jq_iq_j\int_d^\infty\mathrm{d}rr^2v\ce(r)H_{ij}(r),
\ee
where we accounted for the definition of the total correlation function~(\ref{eq42}) together with the global electroneutrality condition~(\ref{eq32}). 

For the applicability of Eq.~(\ref{a40}) to different liquid models, it is noteworthy that this identity is valid for a general pairwise interaction potential $q_iq_jv\ce(r)$. In the specific case of the three dimensional Coulomb potential $v\ce(r)=\ell_{\rm B}/r$, Eq.~(\ref{a40}) reduces to
\be
\label{a4}
\beta E_{\rm ex}=2\pi\ell_{\rm B}\sum_{i=1}^p\sum_{j=1}^pn_in_jq_iq_j\int_d^\infty\mathrm{d}rrH_{ij}(r).
\ee

\subsection{Osmotic pressure}
\label{appr}

The derivation of the osmotic pressure is based on the virial theorem~\cite{Hansen}
\be\label{a5}
\beta P=\sum_{i=1}^pn_i-\frac{1}{3V}\lan\beta S\ran_{\rm G},
\ee
where we defined the sum
\be
\label{a6}
S=\sum_{i=1}^p\sum_{j=1}^{N_i}\br_{ij}\cdot\nabla_{\br_{ij}}\left[E\ce+E\h\right].
\ee
Expressing Eq.~(\ref{a6}) in terms of the total force ${\mathbf F}_{{\rm tot},ij}=-\nabla_{\br_{ij}}\left[E\ce+E\h\right]$ experienced by the ion $j$ of the species $i$, one obtains
\bea
\label{a7}
S&=&-\sum_{i=1}^p\sum_{j=1}^{N_i}\br_{ij}\cdot {\mathbf F}_{{\rm tot},ij}\\
\label{a8}
&=&-\sum_{i=1}^p\sum_{j=1}^{N_i}\sum_{k=1}^p\sum_{l=1}^{N_k}\br_{ij}\cdot {\mathbf F}_{kl,ij}\left(1-\delta_{ik}\delta_{jl}\right)
\eea
Passing from Eq.~(\ref{a7}) to Eq.~(\ref{a8}), we made used of the superposition principle and introduced the force ${\mathbf F}_{kl,ij}$ exerted by the ion $l$ of the species $k$ on the ion $j$ of the species $i$. Next, by permuting the summation indices and using Newton's third law, Eq.~(\ref{a8}) can be rearranged as
\be
\label{a9}
S=-\frac{1}{2}\sum_{i=1}^p\sum_{j=1}^{N_i}\sum_{k=1}^p\sum_{l=1}^{N_k}\left(\br_{ij}-\br_{kl}\right)\cdot {\mathbf F}_{kl,ij}\left(1-\delta_{ik}\delta_{jl}\right).
\ee
Switching now from the force to the potential picture, and introducing Dirac delta functions, Eq.~(\ref{a9}) can be expressed in the form
\begin{widetext}
\bea
\label{a10}
S=\frac{1}{2\beta}\sum_{i=1}^p\sum_{j=1}^{N_i}\sum_{k=1}^p\sum_{l=1}^{N_k}\int\mathrm{d}^3\br\mathrm{d}^3\br'\delta(\br-\br_{ij})\delta(\br'-\br_{kl})\left(1-\delta_{ik}\delta_{jl}\right)||\br-\br'||\left\{v'\h\left(||\br-\br'||\right)+q_iq_jv'\ce\left(||\br-\br'||\right)\right\},\nonumber
\eea
where the prime sign denotes the derivative of the function with respect to its argument. Consequently, based on the definition~(\ref{eq26}) of the pair distribution function, the GC average in Eq.~(\ref{a5}) becomes
\be\label{a11}
\lan\beta S\ran_{\rm G}=\frac{1}{2}\sum_{i=1}^p\sum_{j=1}^pn_in_j\int\mathrm{d}^3\br\mathrm{d}^3\br'||\br-\br'||\left\{v'\h\left(||\br-\br'||\right)+q_iq_jv'\ce\left(||\br-\br'||\right)\right\}g_{ij}(\br-\br').
\ee
\end{widetext}
Plugging Eq.~(\ref{a11}) into the virial identity~(\ref{a5}), and accounting for the translational symmetry governing the bulk liquid together with the HC constraint  $g_{ij}(r)\propto\theta(r-d)$, the osmotic pressure simplifies to
\bea
\label{a12}
\beta P&=&\sum_{i=1}^pn_i\\
&&-\frac{2\pi}{3}\hspace{-.5mm}\sum_{i=1}^p\sum_{j=1}^pn_in_j\hspace{-1mm}\int_d^\infty\hspace{-2mm}\mathrm{d}rr^3\hspace{-.5mm}\left[v'\h(r)\hspace{-.5mm}+\hspace{-.5mm}q_iq_jv'\ce(r)\right]g_{ij}(r)\nonumber.
\eea

In order to simplify the HC potential component of Eq.~(\ref{a12}), we express the HC constraint embodied by the pair distribution function as $g_{ij}(r)=f_{ij}(r)e^{-v\h(r)}$. Thus, the HC contribution to Eq.~(\ref{a12}) becomes
\bea
\label{a13}
&&-\int_d^\infty\mathrm{d}rr^3v'\h(r)g_{ij}(r)=\int_d^\infty\mathrm{d}rr^3f_{ij}(r)\partial_r e^{-v\h(r)}\nonumber\\
&&=\int_d^\infty\mathrm{d}rr^3f_{ij}(r)\partial_r\theta(r-d)=\int_d^\infty\mathrm{d}rr^3f_{ij}(r)\delta(r-d)\nonumber\\
&&=d^3g_{ij}(d^+).
\eea
Consequently, with the use of the global electroneutrality condition~(\ref{eq32}) and the definition of the total correlation function~(\ref{eq42}), one gets
\bea
\label{a13II}
\hspace{-1cm}\beta P&=&\sum_{i=1}^pn_i+\frac{2\pi d^3}{3}\sum_{i=1}^p\sum_{j=1}^pn_in_j\left[H_{ij}(d^+)+1\right]\nonumber\\
&&-\frac{2\pi}{3}\hspace{-.5mm}\sum_{i=1}^p\sum_{j=1}^pn_in_jq_iq_j\hspace{-1mm}\int_d^\infty\hspace{-2mm}\mathrm{d}rr^3v'\ce(r)H_{ij}(r).
\eea
It is again noteworthy that the pressure identity~(\ref{a13II})  is valid for liquids characterized by a general interaction potential of the form $q_iq_jv\ce(r)$. Finally, specifying in Eq.~(\ref{a13II}) the form of the Coulomb potential $v\ce(r)=\ell_{\rm B}/r$, the pressure becomes
\be
\label{a14}
\beta P=\sum_{i=1}^pn_i+\frac{2\pi d^3}{3}\sum_{i=1}^p\sum_{j=1}^pn_in_j\left[H_{ij}(d^+)+1\right]+\frac{1}{3}\beta E_{\rm ex}.
\ee

\section{Proof of the variational identity~(\ref{eq18})}
\label{pr}

In this appendix, we provide an alternative proof of the variational Eq.~(\ref{eq18}). To this aim, we first define the l.h.s. of Eq.~(\ref{eq18}) as
\be
\label{a14II}
S=\hspace{-.5mm}-\hspace{-2.5mm}\sum_{\gamma=\{{\rm s,l}\}}\hspace{-.5mm}\int\hspace{-1mm}\mathrm{d}^3\br\mathrm{d}^3\br'\left[\partial_{\sigma}v_\gamma^{-1}(\br,\br')\right]\hspace{-1mm}\left\{G_\gamma(\br,\br')-v_\gamma(\br,\br')\right\}.
\ee
Then, we plug Eq.~(\ref{eq43}) into the 2PCF~(\ref{eq38}) to obtain
\begin{widetext}
\be
\label{a15}
G_\gamma(\br,\br')=v_\gamma(\br,\br')-\sum_{i=1}^pn_iq_i\int\mathrm{d^3}\br_1\mathrm{d^3}\br_2v_\gamma(\br,\br_1)Q_i(\br_1,\br_2)v_\gamma(\br_2,\br').
\ee
Substituting Eq.~(\ref{a15}) into the r.h.s. of the identity~(\ref{a14II}), one gets
\be
\label{a16}
S=\sum_{\gamma=\{{\rm s,l}\}}\sum_{i=1}^pn_iq_i\int\mathrm{d}^3\br\;\mathrm{d}^3\br'\left[\partial_\sigma v_\gamma^{-1}(\br,\br')\right]\int\mathrm{d}^3\br_1\mathrm{d}^3\br_2v_\gamma(\br,\br_1)Q_i(\br_1,\br_2)v_\gamma(\br_2,\br').
\ee
Then, we note that via the product rule for derivatives, Eq.~(\ref{a16}) can be expressed as
\bea
\label{a17}
S=\sum_{\gamma=\{{\rm s,l}\}}\sum_{i=1}^pn_iq_i\int\mathrm{d}^3\br\mathrm{d}^3\br_1\mathrm{d}^3\br_2v_\gamma(\br,\br_1)Q_i(\br_1,\br_2)\int\mathrm{d}^3\br'\left\{\partial_\sigma \left[v_\gamma^{-1}(\br,\br')v_\gamma(\br',\br_2)\right]-v_\gamma^{-1}(\br,\br')\partial_\sigma \left[v_\gamma(\br',\br_2)\right]\right\}
\eea
Using now the definition~(\ref{eq37}) of the inverse kernel together with the trivial identity $\partial_\sigma\delta^3(\br-\br_2)=0$, one finds that the first term in the curly bracket of Eq.~(\ref{a17}) vanishes. Thus, one obtains
\bea
\label{a18}
S&=&\hspace{-1mm}-\hspace{-1mm}\sum_{\gamma=\{{\rm s,l}\}}\sum_{i=1}^pn_iq_i\int\mathrm{d}^3\br\mathrm{d}^3\br'\mathrm{d}^3\br_1\mathrm{d}^3\br_2v_\gamma^{-1}(\br',\br)v_\gamma(\br,\br_1)Q_i(\br_1,\br_2)\partial_\sigma v_\gamma(\br_2,\br')\\
&=&\hspace{-1mm}-\hspace{-1mm}\sum_{\gamma=\{{\rm s,l}\}}\sum_{i=1}^pn_iq_i\int\mathrm{d}^3\br'\mathrm{d}^3\br_1\mathrm{d}^3\br_2Q_i(\br_1,\br_2)\partial_\sigma v_\gamma(\br_2,\br')
=-\sum_{i=1}^pn_iq_i\int\mathrm{d}^3\br'\mathrm{d}^3\br_1\mathrm{d}^3\br_2Q_i(\br_1,\br_2)\partial_\sigma v\ce(\br_2,\br').\nonumber
\eea
\end{widetext}
Through the second equality of Eq.~(\ref{a18}), we used again the identity~(\ref{eq37}). Then, in order to pass from the third to the fourth term of Eq.~(\ref{a18}), we carried out the sum over the index $\gamma$ and used the constraint~(\ref{eq6}). Finally, considering that the Coulomb potential does not depend on the splitting parameter $\sigma$, i.e. $\partial_\sigma v\ce(\br_2,\br')=0$, one finds that the last term in Eq.~(\ref{a18}) vanishes, i.e. $S=0$. This implies the cancellation of Eq.~(\ref{a14II}) and therefore completes the proof of the variational identity~(\ref{eq18}).

\section{Conversion of experimentally measured concentrations and osmotic coefficients from molal to molar basis}
\label{conv}

In this appendix, we explain the conversion from the molal basis of the experiments to the molar basis of the SCDH formalism. First, in order to relate the molar concentration $n_i$ of the ion species i and its molal concentration $m_i$, we recall their definitions,
\be
\label{a19}
n_i=\frac{N_i}{V_{\rm el}({\rm L})}=10^{-3}\frac{N_i}{V_{\rm el}({\rm m}^3)};\hspace{5mm}m_i=\frac{N_i}{N_{\rm w}M_{\rm w}}.
\ee
The identities in Eq.~(\ref{a19}) involve the number of ions $N_i$ and the electrolyte volume $V_{\rm el}$, the total number of water molecules $N_{\rm w}$, and their molecular mass $M_{\rm w}$.

Next, we express the mass density of the electrolyte as
\be
\label{a21}
\rho_{\el}(n_i)=\frac{1}{V_{\rm el}}\left[N_iM\s+N_{\rm w}M_{\rm w}\right],
\ee
where we took into account the concentration dependence of the mass density, and introduced the atomic mass $M\s$ of the salt. Replacing now the ion and water numbers in Eq.~(\ref{a21}) by the concentrations in Eq.~(\ref{a19}), the mass density of the solution becomes 
\be
\label{a22}
\rho_{\el}(n_i)=10^3n_i\frac{1+m_iM\s}{m_i}.
\ee 
Inverting Eq.~(\ref{a22}), one finally obtains the ionic molar concentration in terms of molality as
\be
\label{a23}
n_i=\frac{10^{-3}m_i}{1+m_iM\s}\rho_{\el}(n_i).
\ee
In Eq.~(\ref{a23}), all quantities are expressed in SI units.

The experimentally established concentration and temperature-dependent electrolyte densities taken from Ref.~\cite{ElDen} are of the form
\bea
\label{a24}
\rho_{\el}(n_i)&=&\rho_{\rm w}+An_i+BTn_i+CT^2n_i^2+Dn_i^{3/2}\nonumber\\
&&+ETn_i^{3/2}+FT^2n_i^{3/2}.
\eea
Eq.~(\ref{a24}) involves the pure solvent density $\rho_{\rm w}$ and the empirical coefficients $\{A,B,C,D,E,F\}$ whose numerical values are provided in Ref.~\cite{ElDen}. For each specific aqueous solution, the molarity of the electrolyte can be now mapped from its molal concentration via the numerical solution of Eqs.~(\ref{a23})-(\ref{a24}). 

Finally, for comparison with the theoretical osmotic coefficient $\phi=\beta P/(2n_i)$, the experimental osmotic coefficient $\phi_m$ measured in the Lewis-Randall system should be converted to the McMillan-Mayer system via the identity~\cite{Exp2}
\be
\label{a25}
\phi=\phi_m(1+m_iM\s)\frac{\rho_{\rm w}}{\rho_{\el}(n_i)}.
\ee

\section{Gavish-Promislow (GP) model of dielectric decrement}
\label{Gav}

We report here the salt-dependent dielectric permittivity formula of the GP model~\cite{Gavish},
\be
\label{a26}
\e_{\rm el}=\ew+\left(\ew-\e_{\rm ms}\right){\mathcal L}\left(\frac{3\alpha n_i}{\e_{\rm w}-\e_{\rm ms}}\right),
\ee
where $\mathcal{L}(x)=\coth x-x^{-1}$ stands for the Langevin function, $\alpha$ is the ionic excess polarizability, and $\e_{\rm ms}$ is the dielectric constant of the molten salt. The corresponding model parameters used in the present article are provided in Table~\ref{tab1}.

\begin{table}[ht]
\caption{Parameters of the GP Model~(\ref{a26}) taken from Ref.~\cite{NetzMC2}$^{(1)}$ and the ones established in the supplemental material of Ref.~\cite{Buyuk2024}$^{(2)}$, as well as the parameters adjusted in Fig.~\ref{figAp}$^{(3)}$ by fitting experimental dielectric decrement data~\cite{ExpDielDec}.}
\begin{tabular}{c c c}
\hline\hline 
Salt&\hspace{5mm}$\e_{\rm ms}$&\hspace{5mm}$\alpha\;\left({\rm M}^{-1}\right)$\\ [1.0 ex] 
\hline
KI$^{(1)}$ &\hspace{5mm} $30.5$&\hspace{5mm}$-14.69$  \\ [1.0 ex] 
\hline
NaCl$^{(1)}$ &\hspace{5mm} $27.9$&\hspace{5mm}$-11.59$   \\ [1.0 ex] 
\hline
KCl$^{(1)}$ &\hspace{5mm} $35.0$&\hspace{5mm}$-10.02$  \\ [1.0 ex] 
\hline
KF$^{(1)}$ &\hspace{5mm} $22.4$&\hspace{5mm}$-12.0$  \\ [1.0 ex] 
\hline
LiCl$^{(2)}$ &\hspace{5mm} $11.0$&\hspace{5mm}$-13.5$  \\ [1.0 ex] 
\hline
RbCl$^{(2)}$ &\hspace{5mm} $27.0$&\hspace{5mm}$-11.0$  \\ [1.0 ex] 
\hline
CsCl$^{(2)}$ &\hspace{5mm} $27.0$&\hspace{5mm}$-10.0$  \\ [1.0 ex] 
\hline
LiNO3$^{(3)}$ &\hspace{5mm} $16.0$&\hspace{5mm}$-15.0$  \\ [1.0 ex] 
\hline
KBr$^{(3)}$ &\hspace{5mm} $24.0$&\hspace{5mm}$-13.0$  \\ [1.0 ex] 
\hline
LiBr$^{(3)}$ &\hspace{5mm} $12.0$&\hspace{5mm}$-15.5$  \\ [1.0 ex] 
\hline\hline
\end{tabular}
\label{tab1}
\end{table}

\begin{figure}
\includegraphics[width=1.\linewidth]{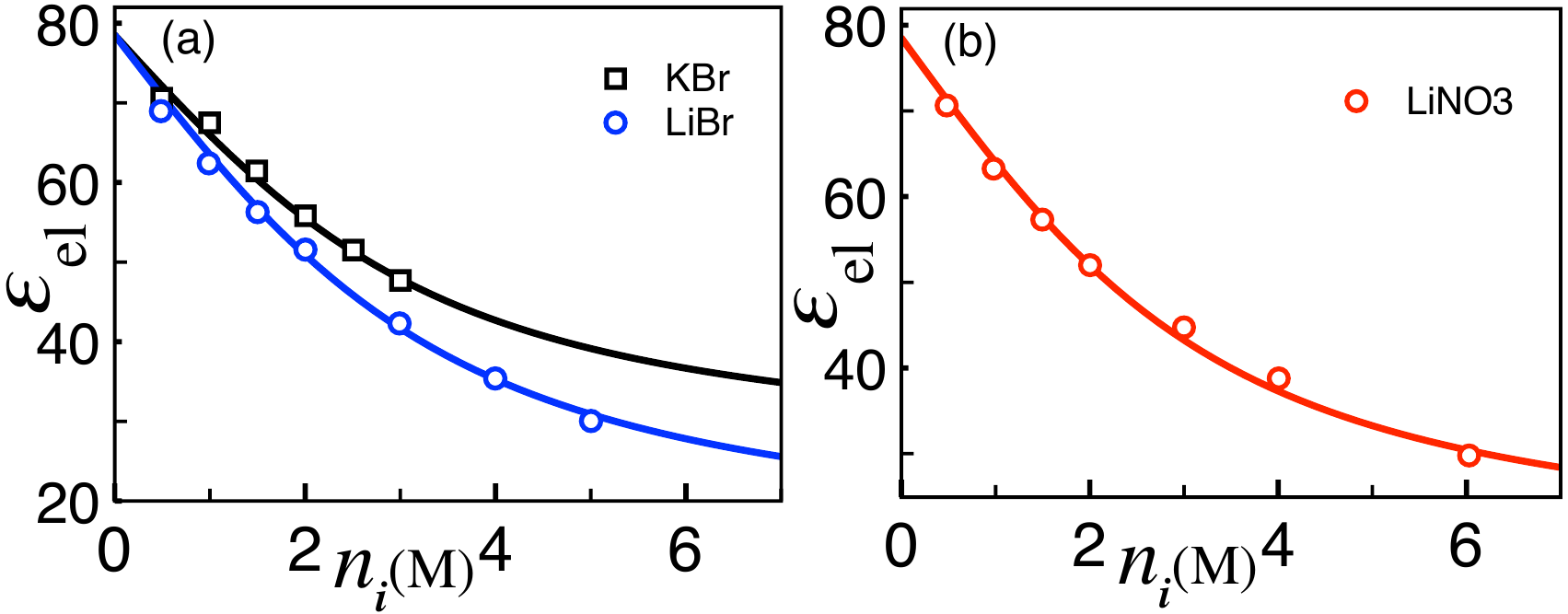}
\caption{(Color online) Dielectric permittivity of the electrolyte against salt concentration. Symbols: experimental data from Ref.~\cite{ExpDielDec}. Solid curves: the prediction~(\ref{a26}) of the GP model~\cite{Gavish}. Table~\ref{tab1} provides the model parameters adjusted to have the best agreement with the experimental data.}
\label{figAp}
\end{figure}

\smallskip
\textbf{Supporting Information}

\smallskip
\textbf{ACKNOWLEDGMENTS}

The author received no financial support for this work.

\newpage
\smallskip
\pagestyle{fancy}
\lhead{TOC Graphic}
\begin{figure}
\includegraphics[width=1.03\linewidth]{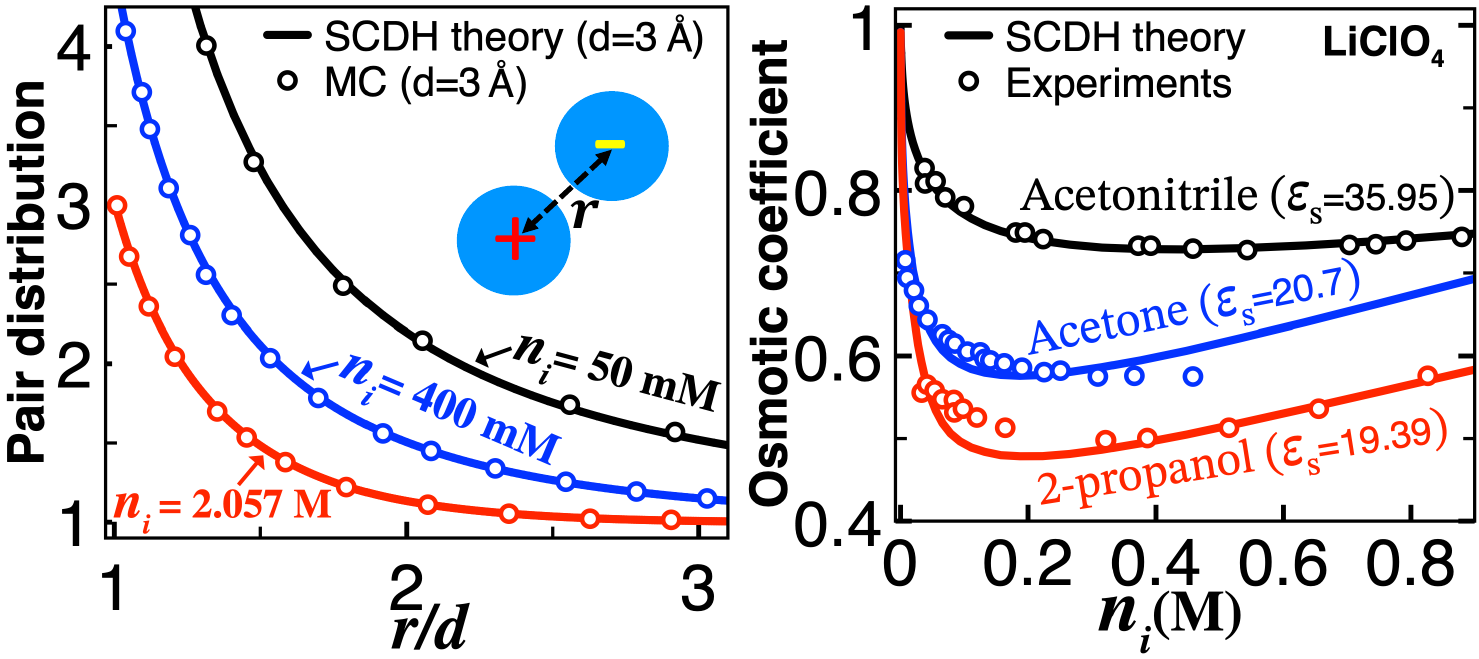}
\caption{TOC Graphic}
\end{figure}


\begin{thebibliography}{99}
\bibitem {rem1} The DH-level pair distributions do not depend on the HC size. However, in Fig.~\ref{fig1}, the ion size affects the scaling of the dimensionless horizontal axis. Thus, the DH predictions of the figure plotted for $d=3$ {\AA} should be compared with the MC and SCDH results corresponding to the same HC size (black curves and symbols).
\bibitem {rem2} In Fig.~\ref{fig5}, the experimental osmotic coefficients and concentrations measured in Ref.~\cite{Exp1} in molal basis have been converted to molar basis via the formulas provided in the Appendix~\ref{conv}.
\bibitem {rem3} To the best of our knowledge,  salt-dependent dielectric permittivity data for the specific non-aqueous solutions in Fig.~\ref{fig8} are not available in the literature. Therefore, in the computation of the SCDH results, the salt-induced variation of the dielectric constant has been neglected. As the osmotic coefficient data in Fig.~\ref{fig8} is limited to submolar concentrations, based on the dielectric decrement curves for aqueous electrolytes (see e.g. Fig.~\ref{figAp}), we expect the corresponding effect to be milder than the case of aqueous solutions considered in Fig.~\ref{fig7}.
\end{thebibliography}
\end{document}